\documentclass[sigconf]{acmart}

\usepackage{graphicx}
\usepackage{enumitem}
\usepackage{tabularx}
\usepackage{makecell}
\usepackage{booktabs}
\usepackage{multirow}
\usepackage{xcolor}
\graphicspath{{img/}}

\newcommand{\paragraphb}[1]{\vspace{0.03in}\noindent{\bf #1} }
\newcommand{\subscript}[2]{$#1 _ #2$}

\begin{document}

\title{Fingerprinting Robot Movements via Acoustic Side Channel}

\author{Ryan Shah}
\affiliation{%
	\institution{University of Strathclyde}
	\country{United Kingdom}
}
\email{ryan.shah@strath.ac.uk}

\author{Mujeeb Ahmed}
\affiliation{%
	\institution{University of Strathclyde}
	\country{United Kingdom}
}
\email{mujeeb.ahmed@strath.ac.uk}

\author{Shishir Nagaraja}
\affiliation{%
	\institution{Newcastle University}
	\country{United Kingdom}
}
\email{shishir.nagaraja@strath.ac.uk}

\begin{abstract}

In this paper, we present an acoustic side channel attack which makes use of
smartphone microphones recording a robot in operation to exploit acoustic
properties of the sound to fingerprint a robot's movements. In this work we consider
the possibility of an insider adversary who is within physical proximity of a
robotic system (such as a technician or robot operator), equipped with only
their smartphone microphone. Through the acoustic side-channel, we demonstrate
that it is indeed possible to fingerprint not only individual robot movements
within 3D space, but also patterns of movements which could lead to inferring
the purpose of the movements (i.e. surgical procedures which a surgical robot
is undertaking) and hence, resulting in potential privacy violations. Upon
evaluation, we find that individual robot movements can be fingerprinted with
around 75\% accuracy, decreasing slightly with more fine-grained movement
meta-data such as distance and speed. Furthermore, workflows could be
reconstructed with around 62\% accuracy as a whole, with more complex movements
such as pick-and-place or packing reconstructed with near perfect accuracy.
As well as this, in some environments such as surgical settings, audio may be
recorded and transmitted over VoIP, such as for education/teaching purposes or
in remote telemedicine. The question here is, can the same attack be successful
even when VoIP communication is employed, and how does packet loss impact the
captured audio and the success of the attack? Using the same characteristics
of acoustic sound for plain audio captured by the smartphone, the attack was
90\% accurate in fingerprinting VoIP samples on average across baseline
movements, which is around 15\% higher than the baseline without the VoIP codec
employed. This is an interesting result as it opens up new research questions
regarding anonymous communications to protect robotic systems from acoustic
side channel attacks via VoIP communication networks.

\end{abstract}


\maketitle

\section{Introduction}
\label{sec:introduction}

The prominence of teleoperated robotic systems has seen a recent rise in a
variety of application areas, such as industrial~\cite{quarta2017experimental} and surgical
environments~\cite{ahn2015healthcare,tewari2002technique}, with promises of higher levels of accuracy and precision.
Given that many of these systems are becoming increasingly connected, they are
vulnerable to an expanded threat landscape in the cyber domain. Attacks from
this angle are primarily active attacks such as tampering with the integrity
of messages in-flight or hijacking the robot controller directly~\cite{bonaci2015make}.
However, little attention has been paid to the capabilities of a passive
attacker and the damage potential of stealthier attacks. Specifically, passive
attacks such side channel attacks which exploit information leakages without the
need to change the normal behaviour of the system, can result in huge losses
that stem from the compromise of operational confidentiality. Side channel
attacks in the cyber domain have the potential to compromise the operational
confidentiality of organisations that own such systems~\cite{shah2022can}, yet those
targeting robots in the physical domain are still to be explored.

In this paper, we aim to investigate whether an adversary can exploit
information leakages from the acoustic side channel, by capturing audible
emanations from a robotic system during normal operations, to mount an attack
that targets operational confidentiality. In this context, we look at two
possible threats posed by an insider attacker. First, a malicious robot operator
or technician on the ground could use a recording device, such as a smartphone,
near the robot to record entire workflows or individual movements. By
fingerprinting this leaked information, they could sell this on to competing
organisations for a malicious advantage. While it can be argued that an attacker
may not be able to get close enough to the robot to place the recording device,
many robotic systems now employ sensors to aid safety mechanisms to prevent harm
to nearby humans or environmental changes. This can allow the attacker windows
of opportunity to place the recording device near the robot or be near enough
to capture meaningful acoustic emanations. A second possible threat comes from
a telemonitoring perspective. While telemonitoring is less common in industrial
settings, in surgical settings the use of medical recording devices, such as
medical data recorders or intraoperative video recorders, are used for
post-surgical review or teaching (alongside patient consent) to learn from
suboptimal scenarios and improve
performance~\cite{saun2019video,5gdigvodafone,pmid32407799}. While privacy laws
and medicolegal requirements govern the use of such devices, data from them is
not typically required as evidence in court so long as patient confidentiality
is maintained~\cite{dalen2019legal}. However, acoustic emanations captured by
such recordings could reveal the operations the robot is carrying out, and
ultimately piece together surgical procedures. In combination with other
metadata, such as patient admission and exit times, this could compromise
patient confidentiality.

In this attack, we recorded the acoustic emanations, through a smartphone
recorder, for individual robot movements, as well as recording entire workflows
corresponding to typical warehousing operations such as picking and placing
objects from one place to another. Using the collected data, we extract a set
of acoustic characteristics which are used as input to an artificial neural
network (ANN). We found that baseline movements (of minimum speed and distance)
can be fingerprinted with at least \textasciitilde75\% accuracy as a
baseline. The speed and distance of movements are not as successfully
fingerprinted in this attack, compared to the radio frequency side channel.
Entire warehousing workflows with \textasciitilde64\% accuracy. Ultimately, it
is clear that a passive insider adversary has the potential not only to reveal
what a robot is doing but take the resulting liabilities of such an attack to
an extreme that impacts even the organisations that employ them. As well as this,
in certain robotics environments, such as in surgical settings, procedures may
be streamed and/or recorded for viewing, education or
research~\cite{muensterer2014google,kulkarni2020cloud,hosseini2013telesurgery}.
Therefore, it is important to question how VoIP impacts the audio samples for
movements and workflows and, ultimately, the success of the attack. Using the
Opus codec -- a common choice for most modern VoIP applications -- the attack
was 90\% accurate for computing movement fingerprints for baseline speed and
distance, which is nearly 15\% more accurate than the baseline without the Opus
codec employed, presenting new research questions regarding side channel attacks
via VoIP communication networks which target robotic systems.

The remainder of this paper is as follows. In Section~\ref{sec:background} we
provide background on teleoperated robots and acoustic emanations, to which we
then describe the threat model. We then outline the attack and our findings in
Section~\ref{sec:attackeval}, and provide an in-depth discussion in
Section~\ref{sec:discussion}. In Section~\ref{sec:related} we discuss related
work and conclude in Section~\ref{sec:conclusion}.

\section{Background}
\label{sec:background}

\subsection{Teleoperated Robots}

\begin{figure}[h]
	\centering
	\includegraphics[width=0.8\linewidth]{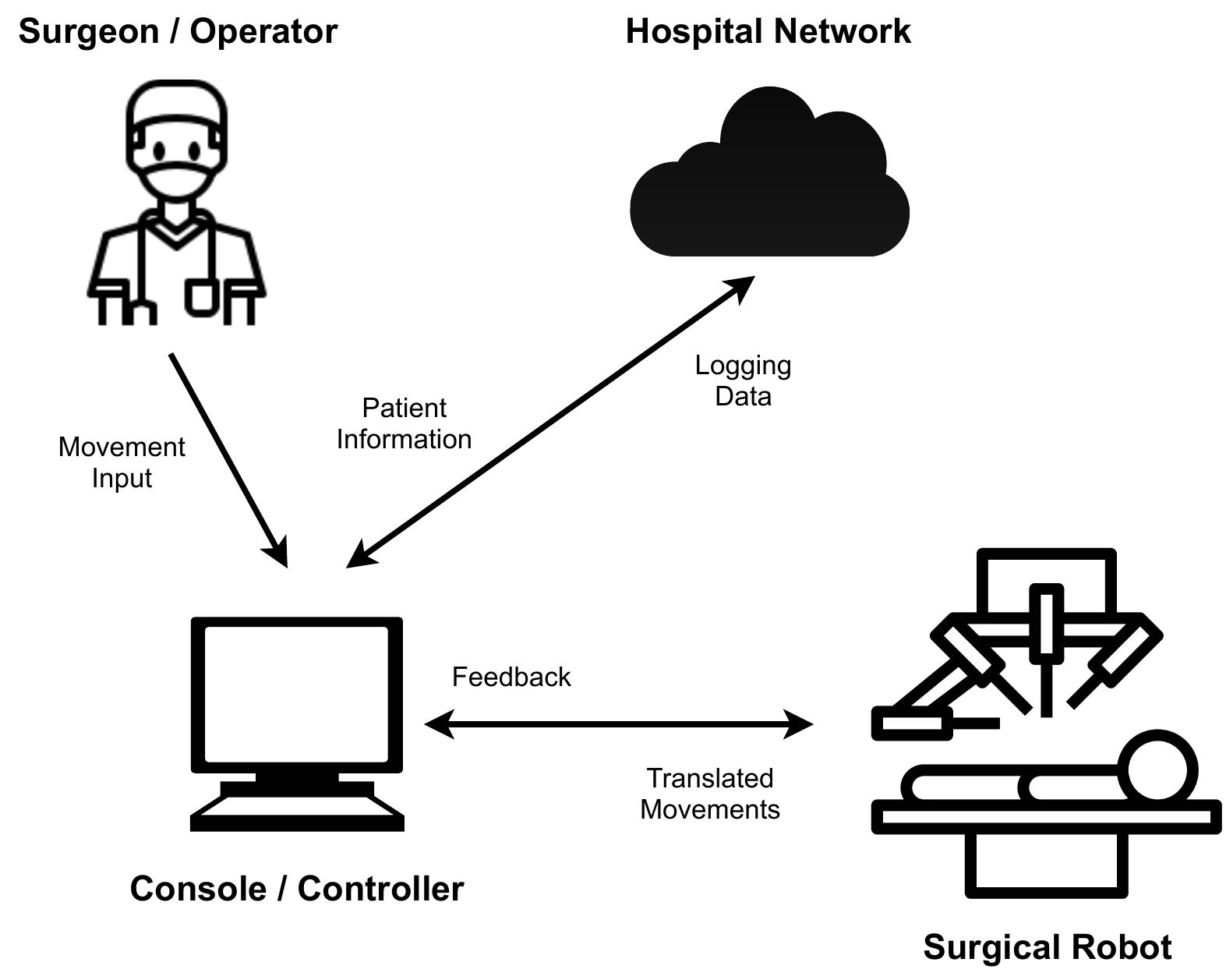}
	\caption{Teleoperated Robot Architecture}
	\label{fig:teleoparch}
\end{figure}

The use of robotics has seen an increase in installations in a variety of
application areas~\cite{reutersrobots} and play pivotal roles bringing benefits
to quality of service, efficiency and precision, among others. Among them,
teleoperated robots are most prominent in many
industrial~\cite{aschenbrenner2015teleoperation,avila2020study,grabowski2021teleoperated,li2017teleoperation,bartovs2021overview}
and surgical~\cite{sung2001robotic,tewari2002technique,hannaford2012raven}
environments, and share a common system architecture. This type of system makes
use of a human operator (i.e. a specially-trained surgeon) who operates the
controller (i.e. surgeon's console or teach pendant) which translates human
movements or inputs into those which the robot can interpret. These input
(console and other sources of information) and output (actuators) devices are
linked together via an electronic control system (ECS) and typically connected
to the organisation's network in which the robot operates. An overview of a
typical teleoperated robot architecture can be seen in
Figure~\ref{fig:teleoparch}.

\subsection{Acoustic Characteristics}

While the robot moves, its electromechanical components will emit (audible)
sound, which when captured could be used to mount an information leakage
attack. The first step to determining an appropriate attack strategy is to
understand the different characteristics of acoustic emanations, and what may
be most useful from an attack perspective.

\subsubsection{Root Mean Square Energy}
Root Mean Square (RMS) energy~\cite{bachu2008separation} is a measure of the
amplitude based on all samples in a frame of audio, and can be thought of as an
indicator of loudness of the audio signal~\cite{panagiotakis2005speech}. This
may be useful in the context of this attack given that combinations of movements
(i.e. simultaneous movements along 2 or more axes) may emanate a louder sound
given the use of more stepper motors, for example. As well as this, as the robot
axes pass the microphone, the sound may be louder and thus this feature may help
provide further information to the discrimination between movements in different
positions.

\subsubsection{Zero-Crossing Rate}
Zero-Crossing Rate (ZCR) is a measure of the number of times a signal crosses
the horizontal time axis and can help identify pitch variations in monophonic
tones (sound emitted from one location)~\cite{panagiotakis2005speech}. Given the
robot is stationary in this case, the ZCR may be a useful feature candidate.

\subsubsection{Spectral Centroid}
The spectral centroid provides information corresponding to frequency bands
contain most of the energy, wherein lower centroid (energy) values is likened
to {\em duller} sounds and higher centroid values for {\em brighter}
sounds~\cite{le2011investigation}. In a robotic system, smaller movement
distances and speeds will naturally require less energy and appear more dull
sounding to the human ear, whereas faster and longer movements have better
tonality, and may ultimately provide useful for distinguishing between different
movements of the same source.

\subsubsection{Spectral Bandwidth}
Spectral bandwidth is defined as the {\em full} width of band of light
(wavelength interval) at half the peak
maximum~\cite{klapuri2007signal,atahan2021music}. Acoustic signals oscillate
about a point, and the bandwidth for each time interval in a signal is the sum
of maximum deviation on both sides of this point. The point of the centroid of
the signal may vary for different robot movements and thus may be an important
feature for fingerprinting.

\subsubsection{Spectral Rolloff}
Spectral rolloff is the fraction of frequency bins under a cutoff point where
the total energy of the spectrum is contained, and can help distinguish between
noisy sounds and more harmonic sounds (below roll-off
point)~\cite{kos2013acoustic}. This feature may provide useful to this attack as
it can roll off frequencies that may fall outside of the {\em useful} range of
frequencies where the energy of the acoustic energy of movements is contained.

\subsubsection{Spectral Contrast}
Spectral contrast is the measure of energy of frequencies in windows of
time~\cite{jiang2002music} and can help identify strong spectral peaks to
reflect the distribution between harmonic and non-harmonic components of the
acoustic emanations. As a robot moves, the frequency contents may have energy
that changes with time and capturing the spectral contrast can help measure this
energy variation.

\subsubsection{Chroma Feature}
Chroma feature, sometimes referred to as a chromagram, profiles a sound into
12 {\em pitch class profiles}~\cite{muller2015fundamentals}. In music analysis,
the attempt is to capture the harmonic and melodic characteristics of a song
where pitches can be categorised to one of the scales in the equally-tempered
set of the notes\\
$\{C, C\#, D, D\#, E, F, F\#, G, G\#, A, A\#, B\}$~\cite{cho2013relative,paulus2010state}.
While recorded robot movements are not akin to songs that are analysed in this
fashion, the pitch of sound may correlate with the speed and distance of
movement and may provide useful as a mid-level feature for fingerprinting
movements.

\subsubsection{Mel-Cepstrum Frequency Coefficients}
The Mel scale is a scale of pitches that is felt to be equal in distance from
one another. For example, in audible acoustics listened by a human, differences
in frequency content can be observed if the source of acoustic emanations are in
the same distance and atmosphere~\cite{greenwood1997mel,martinez2012speaker}.
The short-term power spectrum of acoustic emanations can be represented by the
Mel frequency cepstral (MFC) and a combination of coefficients (MFCCs) make up
the MFC. The MFC equally distributes frequency bands to approximate human
auditory response. If variations in robot movements can be inferred from audible
sound, then looking at the mel frequency coefficients (the list of amplitudes
of the spectrum in the mel scale) will provide useful information to the attack.

\subsection{Threat Model}

Many previous attacks focus on an active attacker, which can involve the
tampering of messages~\cite{bonaci2015make} or replaying attacks between the
robot or controller~\cite{mcclean2013preliminary}. In this work, the primary
attacker is a passive insider, such as a malicious technician or operator. Being
an insider close to the robot would allow them to record the acoustic emanations
during the robot's normal operations using a smartphone, which they may have on
them and use covertly~\cite{peng2013mobile}. As well as this, it is also
possible than an insider attacker is able to covertly {\em plant} a microphone
which could transmit recorded audio to the attacker remotely or be retrieved at
a later time. In either case, if an attacker is able to mount an information
leakage attack to fingerprint robot movement patterns from acoustic emanations,
this could lead to the revelation of daily workflows (i.e. in a warehouse) and
ultimately compromise the operational confidentiality of the organisation. For
example, this information could be given to competitors to gain an advantage or
use it maliciously.

A second possible threat comes from a telemonitoring perspective. While
telemonitoring is less common in industrial settings, in surgical settings the
use of medical recording devices, such as medical data recorders or
intraoperative video recorders, are used for post-surgical review or teaching
(alongside patient consent) to learn from suboptimal scenarios and improve
performance~\cite{saun2019video,5gdigvodafone,pmid32407799}. While privacy laws
and medicolegal requirements govern the use of such devices, data from them is
not typically required as evidence in court so long as patient confidentiality
is maintained~\cite{dalen2019legal}. However, acoustic emanations captured by
such recordings could reveal the operations the robot is carrying out, and
ultimately piece together surgical procedures. In combination with other
metadata, such as patient admission and exit times, this could compromise
patient confidentiality.

Ultimately, reviewing the nature of acoustic emanations in robotic systems, as
well as the proposed threat model, the aim is to investigate whether an advesary
will be able to record the acoustic emanations from a robot during its normal
operation, and make use of distinct features present across the recorded audio
to fingerprint robot movements and workflows. Several hypothetical factors will
come into play which could influence the potential success of this attack.
First, the type of operations being carried out by the robot can vary in terms
of speed and distance of movement, and so the attack should be robust enough to
fingerprint between these parameters. Second, the distance at which the
microphone is situated away from the robot, naturally due to the Doppler
effect~\cite{doppler1903ueber} wherein sounds soften with distance, will also
have an impact on the success of the attack and should be investigated. Finally,
given that in some cases VoIP technology will be employed, such as for recording
purposes or to livestream medical procedures with surgical robots, the impact
of VoIP on the attack should be evaluated.

\subsection{Hypotheses and Goals}
In this work we aim to investigate whether an advesary will be able to
effectively record the acoustic emanations from a robot during its normal
operation, and make use of distinct features present across the recorded audio
to fingerprint robot movements and workflows. We hypothesise that several
factors will come into play which could influence the potential success of this
attack. First, the type of operations being carried out by the robot can vary in
terms of speed and distance of movement, and so the attack should be robust
enough to fingerprint between these parameters. Second, we also hypothesise that
the distance at which the microphone is situated away from the robot, naturally
due to the Doppler effect~\cite{doppler1903ueber} wherein sounds soften with
distance, will also have an impact on the success of the attack and should be
investigated. Ultimately, the following research questions are proposed:

\begin{enumerate}[label=(\subscript{R}{{\arabic*}})]
	\item Can an attacker fingerprint individual robot movements on each axes,
		  as well as permutations of them?
		  \item How is movement fingerprinting affected by:
  	\begin{enumerate}[label=(\roman*)]
  		\item The speed and distance of movements?
  		\item The distance the recording device (i.e. smartphone) is away from the
            robot?
  	\end{enumerate}
  \item Can entire robot workflows be reconstructed from acoustic emanations?
  \item How do VoIP codecs influence the success of the attack?
\end{enumerate}

\section{Attack Methodology}
\label{sec:attackeval}

In this paper, we investigate an acoustic side-channel attack which exploits
audio emanations from a robot during its operation. Specifically, the aim of
this attack is to fingerprint a robots movements from acoustic characteristics
alone, recorded by smartphone devices in a passive manner. For subsequent
discussion, we aim to answer the following questions:

\begin{enumerate}[label=(\subscript{R}{{\arabic*}})]
	\item Can an attacker fingerprint individual robot movements on each axes,
		  as well as permutations of them?
		  \item How is movement fingerprinting affected by:
  	\begin{enumerate}[label=(\roman*)]
  		\item The speed and distance of movements?
  		\item The distance the smartphone or microphone is away from the robot?
  	\end{enumerate}
	\item Can an attacker recover information about the objects a robot is
		  handling, such as its weight?
\end{enumerate}

\begin{figure}[h]
	\centering
	\includegraphics[width=1.0\linewidth]{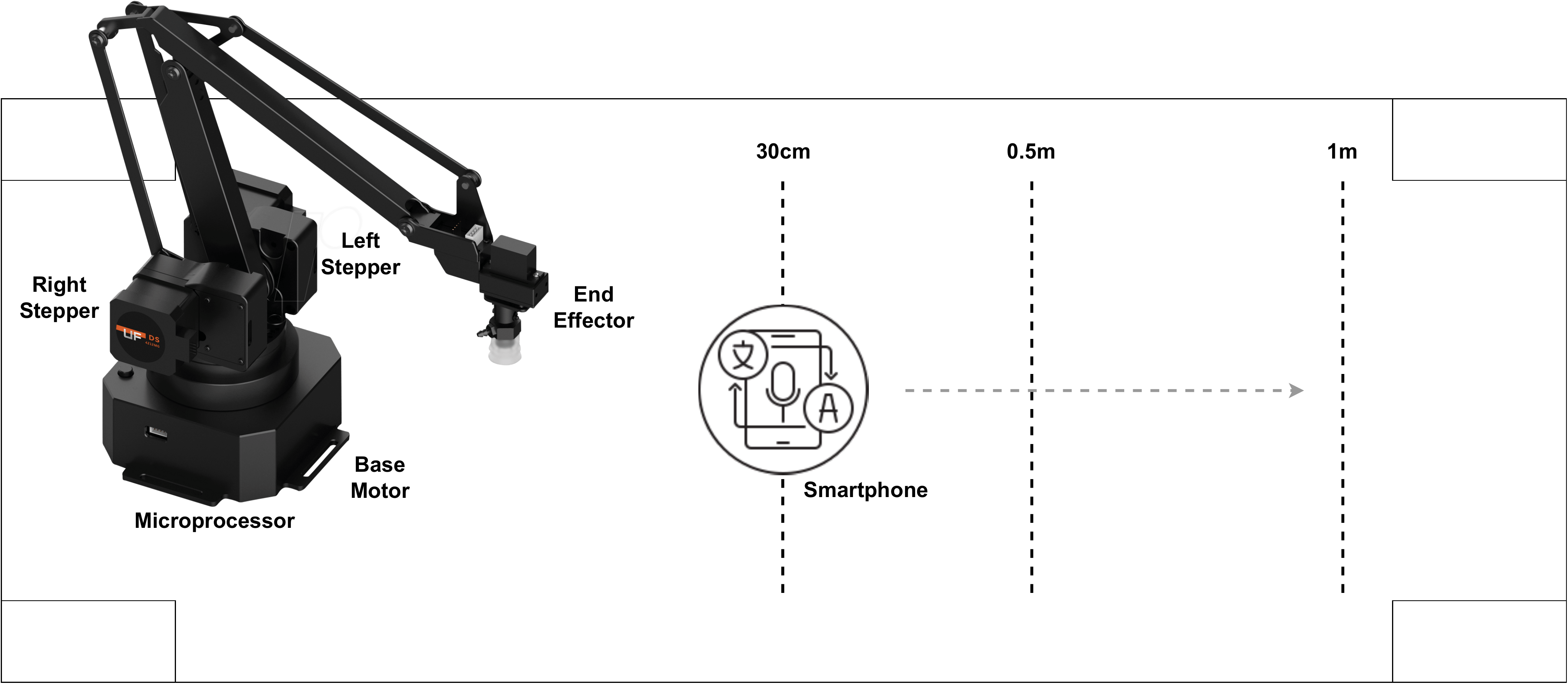}
  \caption{Robot Environment for Acoustic Side Channel}
	\label{fig:robotenv}
\end{figure}

\subsection{Robot Environment}

The context of this study surrounds modern teleoperated surgical robots, whos
typical architecture can be viewed (at a high level) as a pairing between the
robotic system itself and its controller (surgeon's console). For this work, we
use uFactory's uARM Swift Pro which runs on an Arduino Mega 2560 with
MicroPython installed. The controller is emulated on a Windows 10 laptop which
uses the uARM Python (3.8.X) SDK to enable controller instructions to be
written in Python which are then translated into Gcode that is understood by the
robot. An overview of the robot environment used in this study is depicted in
Figure~\ref{fig:robotenv}. For capturing the acoustic emanations which arise
when the robot operates, we position the robot in the center of a table with
the smartphone/microphone placed in several distances away (30cm to 1m) from the
robot as shown in Figure~\ref{fig:robotenv}.

\subsection{Experiment Parameters}

With the robot setup for evaluating our acoustic side-channel attack for
fingerprinting the robot's movements, we now outline the parameters of our
study. Specifically, we will discuss the speed and distance of the movement
operation being carried out, the type of smartphone/micrphone, the distance
the smartphone microphone is away from the robot and finally,

\paragraphb{Speed and Distance.}
In addition to capturing the acoustic emanations which arise during operation
along the X, Y and Z axes, and combined movement operations, it is important
to evaluate more fine-grained movements. To this, we programmed robot
movements with varying distances (in millimetres) as well as varying speeds
of movement (mm/s). This is because in realistic cases, a surgical robot
for example would not move in each direction with constant distance and speed.
Therefore, it is vital to understand whether an adversary can also fingerprint
meta-information as well as just the movements themselves.

\paragraphb{Microphone Distance.}
In terms of recording the acoustic emanations during robot operation, it is
important to evaluate the impact of distance the microphone is away from the
robot. In a real situation, it is highly unlikely that an adversary would be very
close to or in front of the robot, especially in cases like surgical robots where
it could not only be dangerous to stand too close but being close enough may
trigger potential safety features implemented to prevent injury. For this study,
given the size of our uARM robot ($150mm \times 140mm \times 281mm$) and the
volume of sound which is given off during its operation, we cannot investigate
large distances as would be granted with a large surgical robot, for example.
However, given this limitation, we recorded sounds at distances ranging from 30cm
to 1m.

\paragraphb{VoIP.}
The final parameter for this study is to evaluate the impact VoIP has on the
success of the attack. For this study, the codec employed by the majority of
VoIP applications is Opus~\cite{valin2012definition,valin2016high}. The first
step is to observe how the codec performs, but also how packet loss will also
affect audio quality and the success of the attack.

\begin{figure}
	\centering
	\includegraphics[width=1.0\linewidth]{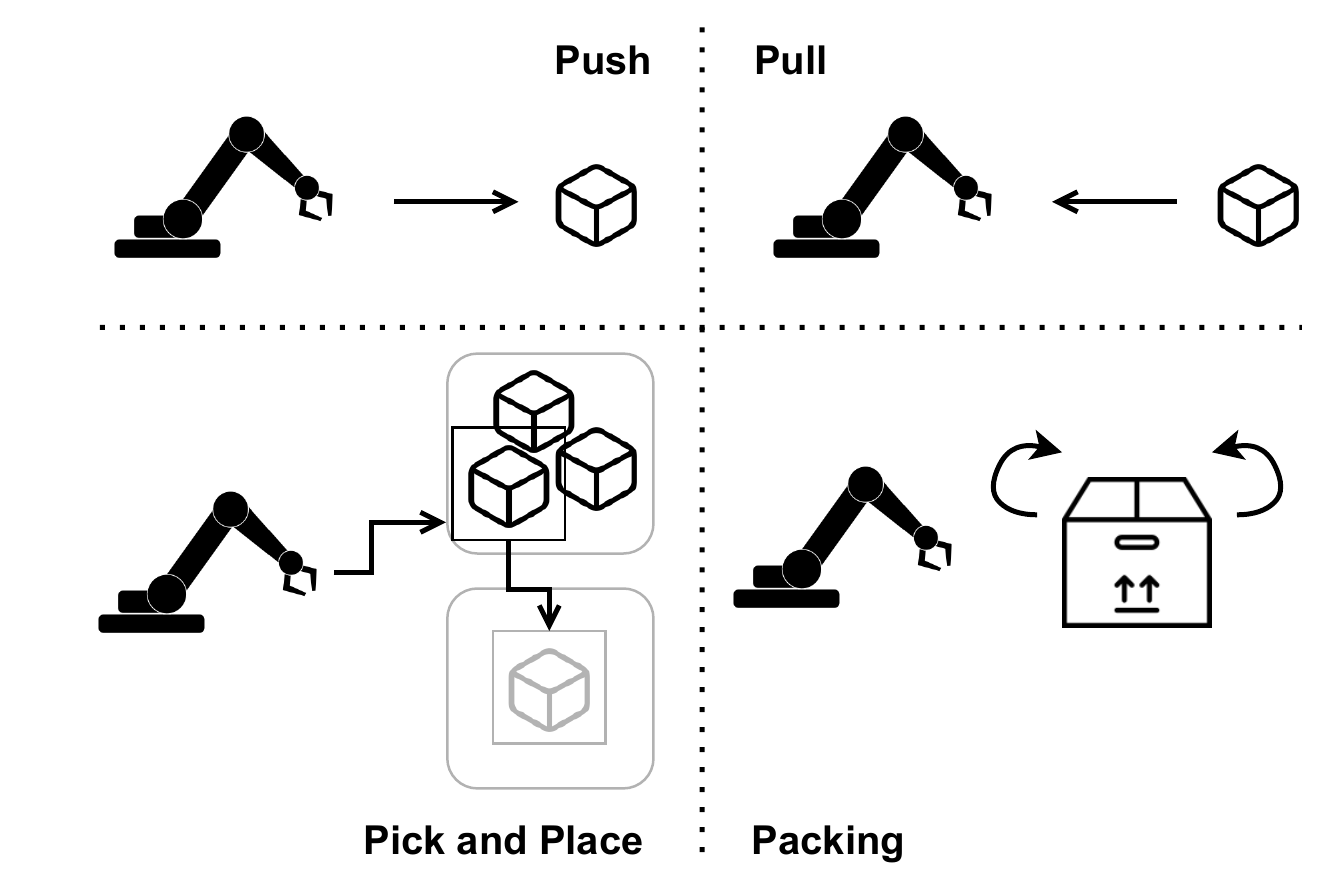}
    \captionsetup{singlelinecheck=off}
    \caption{\centering Depiction of Common Warehousing Workflows\hspace{\textwidth}{\textcolor{darkgray}{\small\textmd{Our dataset contains common warehousing workflows such as pushing, pulling, packing and moving objects}}}}
	\label{fig:workflows}
\end{figure}

\subsection{Movement Dataset}

After determining the appropriate acoustic features to extract from the captured
sounds, the next step was to create the dataset. In this dataset, there are 2
subsets. Within both subsets, there are samples pertaining to both individual
and permutations of movements with varying speeds and distances of movement,
the microphone distance, and robotic warehousing workflows (Figure~\ref{fig:workflows}). These workflows
include those such as pick-and-place, push and pull operations, which were replicated from those
found in existing industrial robot datasets such as the {\em Forward Dynamics
Dataset Using KUKA LWR and Baxter}~\cite{polydoros2016reservoir} for pick and
place and the {\em Inverse Dynamics Dataset Using
KUKA}~\cite{rueckert2017learning} for push/pull. For these workflows, movements
were slightly perturbated to account for a small degree of entropy that may be
present in real-world operations (i.e. those that may arise due to drift in
equipment calibration or wear-and-tear). In contrast to the first subset, the
second subset contains the same samples but are passed through the Opus codec
to evaluate the impact of VoIP on recorded audio in this attack. Specifically,
while all samples are passed through the Opus codec, they are further split by
packet loss. Packet loss has been shown to negatively impact call quality in
VoIP communications~\cite{ortega2018evaluation,laghari2020effect}, as they
induce impact in the form of dropped calls or parts of speech, slow rate of
speech (latency) or noise/interference. Because of this, these further subsets
are divided by packet loss values of 1\%, 5\%, 10\%, 25\% and 50\%. As a whole,
the first subset contains 27.2K samples for individual movements and 658 samples
for warehousing workflows, with each using 20\% of the total samples for
validation and another 20\% for testing. The second contains the same amount of
samples for each of the packet losses evaluated.

\paragraphb{Dataset Pre-Processing.}
The features in the dataset, as listed above, are computed using the
{\em librosa}~\cite{mcfee2015librosa} Python library. For each feature, the mean
value of each feature across each signal sample is taken and computed from a
Short-Time Fourier Transform (STFT) with a Hann window and FFT length of 8192.
For the MFCCs, 14 coefficients were used. Typically, 8--13 are used with the
zeroth excluded given it only represents the average log-energy of the input
signal~\cite{rao2014speech}. However, given this is a new problem to be
explored, this is also kept to later examine its importance for fingerprinting.

\subsection{Neural Network}
Before an evaluation can take place, an important step is constructing an
appropriate neural network architecture for fingerprinting movements and
ensuring a successful attack. To create the neural network, a sequential model
was used where neurons are grouped in a linear fashion. This was created using
the Keras API~\cite{chollet2015keras}. The parameters and structure for the
layers in the neural network were evaluated on the dataset using a
cross-validated grid search to find the most optimal number of neurons, layers,
acttion function and dropouts if necessary. The input for the maximum number of
neurons to be tested was calculated using the formula proposed by Demuth et
al.~\cite{demuth2014neural} with an alpha branching factor of $2$. Using
the grid search with 3 cross validations, the most optimal neural network
architecture for this feature set consists of 5 layers. First, the input layer
containing 21 neurons for each of the input features. Next, there are 4 hidden
layers. The first is a {\em Dense} layer with 290 neurons and uses the ReLU
activation function~\cite{eckle2019comparison}. The next hidden layer is a
{\em Dropout} layer which is used to randomly set input units to 0 at a rate
of $0.05$ at each step during training to prevent overfitting. The next
layer is another {\em Dense} layer of 350 neurons with ReLU activation, followed
by another Dropout with a rate of $0.05$ to prevent overfitting. Finally, the
last layer is a {\em Dense} output layer of 7 neurons, one for each of the
movement classes, and uses the SoftMax activation
function~\cite{dunne1997pairing} to have the output in the range of $[0,1]$ for
use as predicted probabilities. Sparse categorical cross-entropy is used as
labels are integers and not one-hot encoded, for which categorical cross-entropy
would be used~\cite{zhang2018generalized}. The optimiser used is Adam with a
learning rate of $0.001$. This learning rate was chosen as others, such as those
with higher learning rates, resulted in lowered accuracy scores. The model was
fitted with a batch size of 32 and was run for 1000 epochs.

\paragraphb{Choice of Activation and Optimisation Functions.}
The ReLU activation function was chosen over other activation functions, as the
reduced likelihood of vanishing gradient allows for a constant gradient
resulting in faster learning. Further, the sparsity of representations are shown
to be more beneficial than dense representations, as seen in other activations
such as sigmoids~\cite{krizhevsky2012imagenet,li2017convergence,agarap2018deep}.
The softmax activation function, combined with categorical
cross-entropy~\cite{zhang2018generalized} for the loss function, was chosen due
to the simple fact that this is a multi-class classification problem. Simply, a
sample can belong to one of the 7 classes, with each class corresponding to one
of the robot movements. As well as this, the Adam optimiser was an ideal
candidate. It is an extension to the Stochastic Gradient Descent (SGD) method,
based on adaptive estimation of first- and second-order
moments~\cite{kingma2014adam}. Specifically, it allows for the updating of
network weights iteratively based on the training data, and fits best with the
weighted sample sets in opposition to other tried methods such as standard SGD,
RMSProp and SGD + Nesterov Momentum.

\section{Evaluation}

After setting up the robot environment and capturing the acoustic emanations
during various stages of operations, the next step is to evaluate the success of
the attack. As per the research questions listed above, the evaluation of this
attack and related results will be set out in that order.

\subsection{Individual Movement Fingerprints}

The first research question ($R_1$) aims to investigate whether an attacker can
infer individual movements (on each axis) and permutations of these movements
from the recorded audio. To compare this against other parameters, this
experiment is considered as a baseline where the speed and distance of movement
are the lowest possible values (1mm and 12.5mm/s respectively), and no VoIP
codec used. As seen in Table~\ref{table:acbaseline}, an average accuracy of
around 75\% can be observed across all movements, with the YZ movement having
the highest precision among the movements. In comparison with the RF side
channel, there is a clear drop in accuracy of around 20\% but the acoustic side
channel outperforms traffic analysis by around 10\%. Interestingly, Y-involved
movements are better recovered than other movements overall, which was not the
case in the RF side channel (albeit a higher accuracy). This may be due to the
Y-axis moving across the microphone range. Looking at the Z-involved movements,
these are among the lowest. This may be due to the Z axis involving a vertical
movement only and not moving nearer the microphone for better recording.

\begin{table}[t]
\centering
\begin{tabular}{ccc}
	\toprule
	\textbf{Movement} & \textbf{Precision} & \textbf{Recall} \\
	\midrule
	X & 76\% & 81\% \\
	Y & 77\% & 78\% \\
	Z & 61\% & 71\% \\
	XY & 78\% & 80\% \\
	XZ & 68\% & 65\% \\
	YZ & 85\% & 78\% \\
	XYZ & 72\% & 67\% \\
  \midrule
  Accuracy & \multicolumn{2}{c}{75\%} \\
	\bottomrule
\end{tabular}
\captionsetup{singlelinecheck=off}
\caption{\centering Baseline Classification Results\hspace{\textwidth}{\textcolor{darkgray}{\small\textmd{As a whole, the baseline accuracy is 75\% which is fairly good inference accuracy for an attacker. Z-based movements show the lowest precision and recall for fingerprinting, perhaps due to vertical motion and no horizontal spread across the recording device}}}}
\label{table:acbaseline}
\end{table}

\subsection{Impact of Movement Distance}

For the next research question ($R_2$), the evaluation will look into how the
distance and speed ($R_{2(i)}$) of robot movements, and the distance of the
recording device ($R_{2(ii)}$), impact the success of fingerprinting movements
from the acoustic side channel. First, as a robot moves, there is likely to be
more sound that can be recovered as the distance of movement increases. As seen
in Table~\ref{table:acdistance}, an increase by a single distance unit increases
the model accuracy by 1\%, improving Y-involved movement precision by around
10\%. furthermore, the Z movement also gains a slight increase in precision.
Unfortunately, this results in lowered accuracy for the other movements. This
increase in distance results in the sound of movement being held for longer and
may either provide useful for distinguishing variance between movements or even
reduce this variance. To explore this, larger distances of movements are
explored.
At 5mm, there is a drop in accuracy of
around 4\%, with X-involved movements having much higher accuracy. At 10mm, the
accuracy of the model overall decreases significantly to 57\%. Y-involved
movements in this case are much poorly fingerprinted, yet X-involved movements
have a further increase in precision. For the Z movement at this stage, there is
unfortunately a further drop in precision but the recall remains relatively
similar. At 25mm, the accuracy starts to improve by 7\% with the X movement
having similar precision and recall to 10mm, and most other movements have an
increase in both precision and recall. Finally, at 50mm, the accuracy nears that
of the baseline and 2mm, however X-involved movement accuracy is significantly
improved.

\begin{table}
  \centering
  \scriptsize
  \begin{tabular}{lcc|cc|cc|cc|ccc}
    \toprule
	\multicolumn{12}{c}{{D = Distance (mm), P = Precision, R = Recall}} \\
	\midrule
    \multirow{2}{*}{} &
      \multicolumn{2}{c}{D=2} &
      \multicolumn{2}{c}{D=5} &
	  \multicolumn{2}{c}{D=10} &
	  \multicolumn{2}{c}{D=25} &
	  \multicolumn{2}{c}{D=50} & \\
      & {P} & {R} & {P} & {R} & {P} & {R} & {P} & {R} & {P} & {R} \\
      \midrule
    X & 69\% & 71\% & 77\% & 87\% & 85\% & 58\% & 83\% & 70\% & 86\% & 84\% \\
    Y & 88\% & 77\% & 77\% & 79\% & 80\% & 54\% & 66\% & 43\% & 90\% & 88\% \\
    Z & 65\% & 83\% & 64\% & 79\% & 51\% & 81\% & 64\% & 71\% & 83\% & 66\% \\
    XY & 68\% & 60\% & 67\% & 63\% & 63\% & 53\% & 57\% & 65\% & 79\% & 81\% \\
    XZ & 62\% & 57\% & 83\% & 47\% & 67\% & 49\% & 60\% & 61\% & 64\% & 79\% \\
    YZ & 94\% & 94\% & 66\% & 83\% & 37\% & 54\% & 55\% & 57\% & 56\% & 59\% \\
    XYZ & 69\% & 81\% & 76\% & 68\% & 45\% & 52\% & 63\% & 84\% & 64\% & 58\% \\
    \midrule
    Accuracy & \multicolumn{2}{c}{76\%} & \multicolumn{2}{c}{72\%} & \multicolumn{2}{c}{57\%} & \multicolumn{2}{c}{64\%} & \multicolumn{2}{c}{74\%} \\
    \bottomrule
  \end{tabular}
  \captionsetup{singlelinecheck=off}
  \caption{\centering Classification Results With Distance Parameter\hspace{\textwidth}{\textcolor{darkgray}{\small\textmd{At a slight increase in distance, the accuracy remains similar to the baseline, but further increases in distances lead to a reduction in fingerprinting accuracy. Notably, unlike the baseline, X-involved movement are better fingerprinted at distance}}}}
  \label{table:acdistance}
\end{table}

\subsection{Impact of Movement Speed}

After looking at movement distance, the next parameter for robot movements is
the speed at which the robot is moving along each of the axes ($R_{2(i)}$). As
seen in Table~\ref{table:acspeed}, the speed parameter is less accurately
fingerprinted by the attack compared to the distance parameter by at least 10\%
on average. Interestingly, a similar pattern is observed rgarding X-involved
movements, with accuracy increasing with speed, except from the XYZ movement.
While there are slight drops in accuracy, the precision and recall across most
movements remains similar as speed increases. This is interesting, as the
initial hypothesis was that a higher speed would result in higher pitched
acoustic emanations, however the results seem to contradict this. In any case,
perhaps the perceptual characteristics for human audio, while a clear pitch
change is present listening to the robot in the lab, the feature algorithms
regarding pitch (i.e. chroma feature) may not pick up on this for robot sounds.

\begin{table}
  \centering
  \scriptsize
  \begin{tabular}{lcc|cc|cc|ccc}
    \toprule
	\multicolumn{9}{c}{{S = Speed (mm/s), P = Precision, R = Recall}} \\
	\midrule
    \multirow{2}{*}{} &
      \multicolumn{2}{c}{S=25} &
      \multicolumn{2}{c}{S=50} &
	  \multicolumn{2}{c}{S=75} &
	  \multicolumn{2}{c}{S=100} & \\
      & {P} & {R} & {P} & {R} & {P} & {R} & {P} & {R} \\
      \midrule
    X & 57\% & 81\% & 54\% & 74\% & 78\% & 53\% & 72\% & 81\% \\
    Y & 79\% & 76\% & 61\% & 42\% & 59\% & 45\% & 72\% & 69\% \\
    Z & 50\% & 56\% & 52\% & 58\% & 62\% & 84\% & 77\% & 75\% \\
    XY & 73\% & 72\% & 46\% & 40\% & 67\% & 57\% & 57\% & 70\% \\
    XZ & 79\% & 57\% & 75\% & 79\% & 57\% & 60\% & 60\% & 56\% \\
    YZ & 67\% & 59\% & 66\% & 45\% & 53\% & 65\% & 66\% & 69\% \\
    XYZ & 65\% & 63\% & 51\% & 66\% & 54\% & 57\% & 62\% & 47\% \\
    \midrule
    Accuracy & \multicolumn{2}{c}{66\%} & \multicolumn{2}{c}{58\%} & \multicolumn{2}{c}{60\%} & \multicolumn{2}{c}{66\%} \\
    \bottomrule
  \end{tabular}
  \captionsetup{singlelinecheck=off}
  \caption{\centering Classification Results With Speed Parameter\hspace{\textwidth}{\textcolor{darkgray}{\small\textmd{The speed parameter performs worse than the distance parameter in the acoustic side channel, a similar pattern as seen with the radio frequency side channel}}}}
  \label{table:acspeed}
\end{table}

\subsection{Microphone Distance}
\label{sec:micdist}

While observing more fine-grained information leakage is useful to an attacker,
one problem that may impact the success of the attack is the distance the
recording device is away from the robot -- in this case, the smartphone. Naturally,
due to the Doppler effect, the intensity of sound (i.e. loudness) decreases over
distances, and one would hypothesise that because of this the accuracy may be
significantly impacted as the distance of recording increases. In this experiment,
two other microphone distances (50cm and 100cm) are also tested in addition to
the baseline recorded at 30cm. While these are not large recording distances,
given the small scale of the robot used for the evaluation of the attack, these
are relatively suitable candidates to be tested. As seen in Table~\ref{table:acmicdist},
as the distance the microphone is away the robot is increased, the accuracy
of the attack compared to the baseline decreases by around 10\% at each
recording distance step. Notably, this is much more significant for Z-based
movements which were previously described to have poorer fingerprinting accuracy
due to the limited range of motion that does not cross the recording device
(remains stationary and moves vertically). In this case, a point a future work
may be to evaluate the impact on position of the smartphone around the robot,
aside from facing in front. Collectively, inference from multiple angles may
provide better fingerprinting accuracy in all cases.

\begin{table}
  \centering
  \scriptsize
  \begin{tabular}{lcc|cc|cc}
    \toprule
	\multicolumn{7}{c}{{MD = Microphone Distance (cm), P = Precision, R = Recall}} \\
	\midrule
    \multirow{1}{*}{} &
      \multicolumn{2}{c}{MD=30} &
      \multicolumn{2}{c}{MD=50} &
	  \multicolumn{2}{c}{MD=100} \\
      & {P} & {R} & {P} & {R} & {P} & {R} \\
      \midrule
			X & 76\% & 81\% & 57\% & 79\% & 75\% & 76\% \\
			Y & 77\% & 78\% & 67\% & 74\% & 67\% & 68\% \\
			Z & 61\% & 71\% & 48\% & 66\% & 45\% & 63\% \\
			XY & 78\% & 80\% & 88\% & 91\% & 64\% & 68\% \\
			XZ & 68\% & 65\% & 61\% & 52\% & 51\% & 40\% \\
			YZ & 85\% & 78\% & 83\% & 54\% & 47\% & 35\% \\
			XYZ & 72\% & 67\% & 52\% & 39\% & 33\% & 33\% \\
    \midrule
    Accuracy & \multicolumn{2}{c}{75\%} & \multicolumn{2}{c}{65\%} & \multicolumn{2}{c}{54\%} \\
    \bottomrule
  \end{tabular}
  \captionsetup{singlelinecheck=off}
  \caption{\centering Classification Results With Microphone Distance\hspace{\textwidth}{\textcolor{darkgray}{\small\textmd{As the microphone distance increases away from the robot being recorded, on average the accuracy decreases around 10\% at each step compared to the baseline - more significantly for Z-based movements}}}}
  \label{table:acmicdist}
\end{table}

\subsection{Workflow Recovery}

The next step in the evaluation looks at whether entire warehousing workflows
can be reconstructed through the acoustic side channel attack. While a pattern
matching approach can be successful using individual movement fingerprints, the
ability to reconstruct entire workflows may be useful from an auditing
perspective, for example, where offsets in {\em normal} movement signals can
be flagged and investigated further. As seen in
Table~\ref{table:acworkflowrecovery}, the explored warehousing workflows can be
recovered on average with around 62\% accuracy. Notably, the pick-and-place and
packing workflows are recovered with much higher success than the push and
pull workflows. Simply, the former have much more variation in the pattern of
movements and thus the variance helps with fingerprinting. In the case of push
and pull movements, they are highly similar and it can be hypothesised that only
the direction of movement away from the microphone (i.e. pull is a reverse of
push) provides at least some degree of accuracy between the two.

\begin{table}[h]
\centering
\begin{tabular}{ccc} 
\toprule
\textbf{Workflow} & \textbf{Precision} & \textbf{Recall} \\
\midrule
Push & 37\% & 16\% \\
Pull & 31\% & 59\% \\
Pick-and-Place & 100\% & 96\% \\
Packing & 97\% & 100\% \\
\midrule
\textbf{Accuracy} & \multicolumn{2}{c}{64\%} \\
\bottomrule
\end{tabular}
\captionsetup{singlelinecheck=off}
\caption{\centering Workflow Reconstruction Results\hspace{\textwidth}{\textcolor{darkgray}{\small\textmd{Common warehousing workflows can be reconstructed in their entirety are better recovered through the acoustic side channel if they are more complex and varied. Push and pull operations are less accurate due to the fact they are very similar movements}}}}
\label{table:acworkflowrecovery}
\end{table}

% ...

\subsection{Impact of VoIP}

In certain robotics environments, such as in surgical settings, procedures may
be streamed and/or recorded for viewing, education or
research~\cite{muensterer2014google,kulkarni2020cloud,hosseini2013telesurgery}.
Therefore, it is important to question how VoIP impacts the audio samples for
movements and workflows and, ultimately, the success of the attack. In many
modern VoIP applications, the Opus codec is the preferred
choice~\cite{valin2012definition,valin2016high} given its standardisation and
rank of higher quality compared to other audio formats for a variety of
bitrates. To explore this,the open-source nature of Opus allows for easy
implementation to encode and decode the audio samples and, during decoding,
investigate various packet losses. In VoIP applications, Packet Loss Concealment
(PLC) is used as a decoder feature for receiving data from an unreliable source,
which masks the effects of packet loss in VoIP communications. In realistic
settings, packets may arrive late, be dropped or be corrupted, which may result
in not only a lowered audio quality but in the worst case, dropped parts of the
audio or the entire audio sample entirely. Given that in VoIP applications, a
1\% packet loss is considered an acceptable rate for VoIP to minimise impact on
call quality~\cite{james2004implementing,amirzade2020reliability}, however in
the event of network failures or availability attacks this may be higher. For
completeness, 5 packet losses of 1\%, 5\%, 10\%, 25\% and 50\% are evaluated.
Furthermore, as it was shown that constant bitrate quality does not perform as
well as variable bitrate quality~\cite{ramo2011voice}, samples are encoded and
decoded with variable bitrate. This experiment used the same model as the
previous experiments, but with a batch size of 256 and 100 epochs of training.
As seen in Table~\ref{table:acvoipbaseline}, the results for the baseline speed and
distance of movement (12.5mm/s and 1mm respectively) under various packet
losses via the Opus codec can be seen. Interestingly, at low packet loss, the
classification accuracy is around 90\% and increases by around 15\% compared to the baseline without
VoIP employed. Further, X movements are more accurately fingerprinted across
all packet losses compared to the baseline without VoIP. As the packet loss
reaches more undesirable amounts of 25\% and 50\%, the accuracy slightly decreases
but the accuracy still remains much higher than the baseline without VoIP.
This may be due to the PLC algorithm switching between CELT or SILK mode and
variable bit rate. Specifically, frames that are deemed important are re-encoded
at a lower bitrate and allows for partial recovery for improtant lost packets.
This may be targetting the movement audio within the sample thus leading to
higher variance among classes.

\begin{table}
  \centering
  \scriptsize
  \begin{tabular}{lcc|cc|cc|cc|ccc}
    \toprule
	\multicolumn{11}{c}{{L = Loss (\%), P = Precision, R = Recall}} \\
	\midrule
    \multirow{2}{*}{} &
      \multicolumn{2}{c}{L=1} &
      \multicolumn{2}{c}{L=5} &
	  \multicolumn{2}{c}{L=10} &
	  \multicolumn{2}{c}{L=25} &
    \multicolumn{2}{c}{L=50} & \\
      & {P} & {R} & {P} & {R} & {P} & {R} & {P} & {R} & {P} & {R} \\
      \midrule
    X & 99\% & 99\% & 99\% & 100\% & 100\% & 100\% & 100\% & 100\% & 99\% & 100\% \\
    Y & 90\% & 94\% & 87\% & 96\% & 90\% & 92\% & 82\% & 97\% & 86\% & 97\% \\
    Z & 86\% & 72\% & 88\% & 68\% & 91\% & 73\% & 86\% & 72\% & 90\% & 74\% \\
    XY & 88\% & 91\% & 91\% & 88\% & 88\% & 91\% & 94\% & 81\% & 90\% & 81\% \\
    XZ & 89\% & 93\% & 82\% & 83\% & 82\% & 82\% & 80\% & 85\% & 80\% & 85\% \\
    YZ & 86\% & 89\% & 87\% & 88\% & 85\% & 89\% & 91\% & 80\% & 91\% & 85\% \\
    XYZ & 93\% & 98\% & 94\% & 97\% & 92\% & 96\% & 89\% & 97\% & 90\% & 97\% \\
    \midrule
    Accuracy & \multicolumn{2}{c}{90\%} & \multicolumn{2}{c}{90\%} & \multicolumn{2}{c}{90\%} & \multicolumn{2}{c}{88\%} & \multicolumn{2}{c}{89\%} \\
    \bottomrule
  \end{tabular}
  \captionsetup{singlelinecheck=off}
  \caption{\centering Classification Results (Baseline) With Opus Codec and Packet Loss\hspace{\textwidth}{\textcolor{darkgray}{\small\textmd{Interestingly, the precision and recall remains relatively similar across packet losses, with a slightly drop in accuracy for undesirable large packet losses. Notably, there is an increase in accuracy of around 15\% compared to the baseline without the Opus codec employed}}}}
  \label{table:acvoipbaseline}
\end{table}

\section{Discussion}
\label{sec:discussion}

The acoustic side channel attack we propose showcases the potential for successfully
compromising the operational confidentiality of organisations in which robotic
systems under attack are deployed. While many active attacks have shown to result
in potentially devastating consequences, the capabilities of a passive insider
attacker are truly underestimated. In this section, a discussion on the proposed
attack is provided.

\subsection{Influence of Noise}

During the recording of acoustic samples for robot movements, there is likely
some degree of background noise that should be accounted for. Given the
recordings were made in a computer lab, background noise effects may include the
likes of light chatter, keyboard tapping and rolling chairs, among others. While
relatively good accuracy is observed even with the background noise, it is
important to also look into techniques to eliminate such noise to determine
whether this results in better fingerprinting accuracy.

In human audio, sound recordings contain the relative signal of the oscillations
due to density and pressure of air in the ear. In digital audio, sound waves are
encoded in digital form as numerical samples in a continuous sequence. The
recordings taken in this attack are recorded at a sampling rate of 44.1KHz with
16-bit depth and thus there are $65,536$ possible values the signal can take in
the sequence.

\begin{figure}[h]
	\centering
	\includegraphics[width=1.0\linewidth]{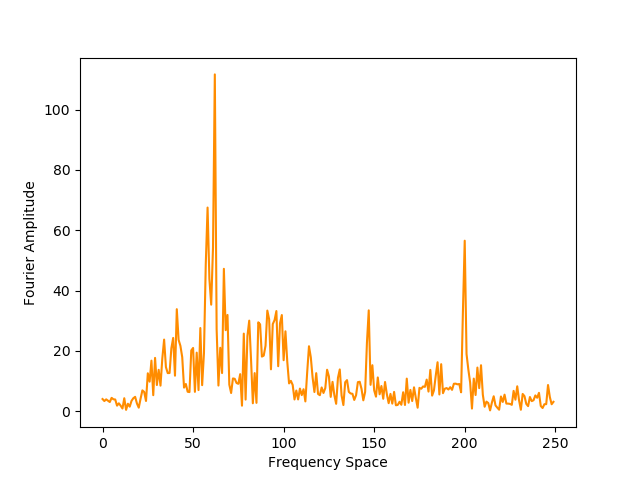}
    \captionsetup{singlelinecheck=off}
    \caption{\centering FFT of Acoustic Signal\hspace{\textwidth}{\textcolor{darkgray}{\small\textmd{Peaks can be observed at 60Hz corresponding to electric hum, with other peaks at 150Hz and 200Hz (among others) which may correlate with robot movement}}}}
	\label{fig:acfft}
\end{figure}
\begin{figure}[h]
	\centering
	\includegraphics[width=1.0\linewidth]{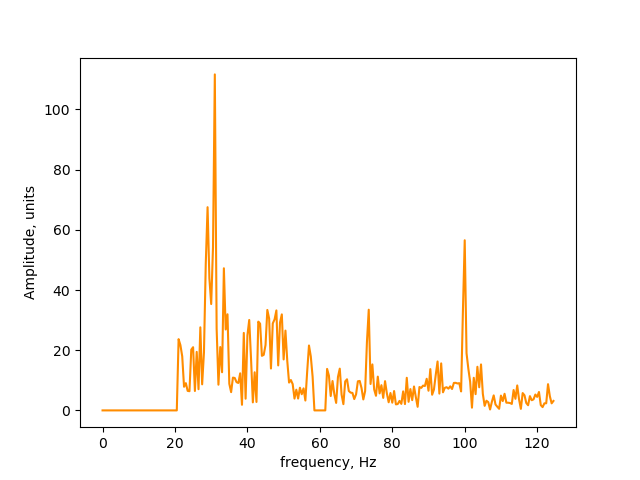}
    \captionsetup{singlelinecheck=off}
    \caption{\centering FFT of Acoustic Signal (Filtered)\hspace{\textwidth}{\textcolor{darkgray}{\small\textmd{The amplitude at points correlating with electric hum or those outwidth the human hearing range are set to 0 (filtered out)}}}}
	\label{fig:acfftfiltered}
\end{figure}

As shown in Figure~\ref{fig:acfft}, the amplitude of the frequency content of
the acoustic signal can be observed using the Fast Fourier Transform (FFT). In
this attack, we make use of techniques originally applied to human acoustics, but
given that the robot movements produce sound that is audible to the human ear
as well. Looking at the frequency content, notable amplitude is not found past
1KHz, so this is zoomed in further to 250Hz. There is a notable spike around
60Hz, which is the frequency standard common to alternating current and is
an effect known as {\em electric hum} due to electrical noise getting into
an acoustic recording medium. The next largest peaks can be observed at around
150Hz and 200Hz which may correspond with the robot movements. As a first step
to noise reduction/filtering, one technique is amplitude filtering, where the
amplitudes of FFT values to be filtered can be set to 0Hz, to which the original
signal can be recreated using an inverse FFT. In this experiment, the electric
hum, as well as frequencies outwidth the human hearing range of
\textasciitilde20Hz--20KHz are filtered by dropping the amplitude of these
ranges. A depiction of the amplitude drop can be seen in
Figure~\ref{fig:acfftfiltered}. Looking at Table~\ref{table:acbaselinefiltered},
the accuracy of baseline movement fingerprints can be observed with amplitude
filtering in place. While the accuracy overall decreases by 1\% compared to the
baseline without amplitude filtering, the precision for Y and XY movements
increase. This may be due to unfortunate noise events present in these samples
that the filter has rectified. However, there is still a reduction in overall
accuracy, which may mean that electric hum and other peaks may not be the best
indicators of noise to remove when recording a robotics system. In this case,
as a point of future work other noise reduction techniques that have shown
to be successful in other areas, such as stationary or non-stationary spectral
gating~\cite{neumann1991spectral,inouye2014towards} which reduces noise in
time-domain signals by estimating noise thresholds for the frequency bands in
a signal to gate (mask) noise below the threshold, are worth exploring in the
hope the attack accuracy may increase.

\begin{table}[t]
\centering
\begin{tabular}{ccc}
	\toprule
	\textbf{Movement} & \textbf{Precision} & \textbf{Recall} \\
	\midrule
	X & 72\% & 81\% \\
	Y & 75\% & 73\% \\
	Z & 68\% & 72\% \\
	XY & 86\% & 71\% \\
	XZ & 69\% & 68\% \\
	YZ & 76\% & 83\% \\
	XYZ & 72\% & 70\% \\
  \midrule
  Accuracy & \multicolumn{2}{c}{74\%} \\
	\bottomrule
\end{tabular}
\captionsetup{singlelinecheck=off}
\caption{\centering Amplitude Filtering Classification Results\hspace{\textwidth}{\textcolor{darkgray}{\small\textmd{While the accuracy is slightly reduced compared to the baseline with no filtering, the precision for some movements increases further, with better recall seen in most cases}}}}
\label{table:acbaselinefiltered}
\end{table}

\subsection{Other VoIP Codecs}

Opus is the primary choice for many VoIP applications due to its royalty free
and open source nature, alongside the benefits of higher quality and
low-bandwidth streaming, in comparison with other codecs such as
Speex~\cite{valin2016speex} or SILK~\cite{vos2010silk} (Opus' predecessor).
While it may be interesting to evaluate other codecs, Opus is the main choice
for the majority of modern applications, such as Zoom, Teams and
Discord~\cite{rajaratnam2018isolated,castro2020your} and is taking over
previously dominating codecs.

\subsection{Defences}
While the attack is successful, and even more so when the attack targets VoIP
communiations, a natural question pertains to countermeasures and defences
against the acoustic side channel attack. In this work, acoustic emanations
result in unintentional information leakages about robot behaviours and can
ultimately lead to the compromise of operational confidentiality.

One defence that could be considered is to make use of vibration- or
sound-reduction mechanisms to hinder the effect of the attack. As seen in
Section~\ref{sec:micdist}, as the microphone distance increases the accuracy
of fingerprinting also decreases. While this is due to the Doppler effect that
is naturally at play with regard to sound intensity (i.e. loudness), a reduction
in this from other means may result in the same outcome of reduced success of
fingerprinting. Techniques in this space include the likes of using vibration
isolation pads~\cite{desaiexperimental} or damping to reduce
vibration~\cite{gravagne2001good,khan2020sliding} for the robot as a whole. In
the case of noise reduction for robot components such as stepper motors,
potential defences include using a clean damper~\cite{ma2019practical} or higher
resolution stepper motors.

Another potential defence is to make use of a masking noise, to interfere with
attack inference by distorting the signal related to information leakage in the
acoustic side channel~\cite{backes2010acoustic,anand2016sound,kim2015vibration}.
Adding a masking signal has shown success, but two challenges need to be
addressed. First, the mask must be similar to the signal requiring masking to
ensure difficult separation. Second, the masking noise should not cause any
degrading effect on usability of the robotic system. For example, if the masking
noise is to cover up other sound such as those used for emergencies or other
operator feedback, then this will be much less than ideal and potentially lead
to catastrophic liabilities.

\section{Related Work}
\label{sec:related}

While acoustic side channel attacks have not been explored for robotic systems,
enhancing the novelty of this work, there has been previous research in the
area of acoustic side channels.
In a similar respect to robotics, the exploration of information leakage in the
acoustic side channel has been explored for 3D printers~\cite{backes2010acoustic}
-- some of which making use of smartphones to carry out the
attack~\cite{song2016my,bilal2017review} -- and additive manufacturing
systems~\cite{chhetri2017confidentiality}. However, many of these attacks solely
focus on IP theft. The acoustic side channel attack presented in this work
focus solely on the movement of the robot arm and the compromise of operational
confidentiality, which when looking at the bigger picture is much more valuable
to an attacker. Furthermore, the reconstruction of G-code is an unnecessary
extra step as movements which correspond to these can be inferred from individual
movement fingerprinting under the assumption the robot is operated by an Arduino.
Furthermore, while the robot in this work is operated by an Arduino, the focus
is on reconstructing movements from the acoustic emanations, irrespective of the
microcontroller used and thus applies to robotic systems in general and not
those restricted to being operated by an Arduino.

\section{Conclusion}
\label{sec:conclusion}

In conclusion, it is clear that even acoustic emanations provide a high level of
accuracy for fingerprinting movements and showcases a highly important passive
side channel attack in the physical domain, which can be carried out with a
fairly cheap smartphone. While more fine-grained movements and entire workflows
in warehousing settings can be inferred, our contributions demonstrate that
the recent usage of VoIP technologies also leave potential for information
leakage through these communication channels, with the result leaving movement
fingeprints to be more accurately reconstructed. This is an interesting result
as it opens up new research questions regarding anonymous communications to
protect robotic systems from acoustic side channel attacks via VoIP
communication networks.

\begin{acks}
	The authors are grateful for the support by the Engineering and Physical Sciences
	Research Council (11288S170484-102) and the support of the
	National Measurement System of the UK Department of Business, Energy \&
	Industrial Strategy, which funded this work as part of NPL's Data Science program.
\end{acks}

\bibliographystyle{ACM-Reference-Format}
\bibliography{references}


\begin{thebibliography}{71}


\ifx \showCODEN    \undefined \def \showCODEN     #1{\unskip}     \fi
\ifx \showDOI      \undefined \def \showDOI       #1{#1}\fi
\ifx \showISBNx    \undefined \def \showISBNx     #1{\unskip}     \fi
\ifx \showISBNxiii \undefined \def \showISBNxiii  #1{\unskip}     \fi
\ifx \showISSN     \undefined \def \showISSN      #1{\unskip}     \fi
\ifx \showLCCN     \undefined \def \showLCCN      #1{\unskip}     \fi
\ifx \shownote     \undefined \def \shownote      #1{#1}          \fi
\ifx \showarticletitle \undefined \def \showarticletitle #1{#1}   \fi
\ifx \showURL      \undefined \def \showURL       {\relax}        \fi
\providecommand\bibfield[2]{#2}
\providecommand\bibinfo[2]{#2}
\providecommand\natexlab[1]{#1}
\providecommand\showeprint[2][]{arXiv:#2}

\bibitem[\protect\citeauthoryear{Agarap}{Agarap}{2018}]%
        {agarap2018deep}
\bibfield{author}{\bibinfo{person}{Abien~Fred Agarap}.}
  \bibinfo{year}{2018}\natexlab{}.
\newblock \showarticletitle{Deep learning using rectified linear units (relu)}.
\newblock \bibinfo{journal}{\emph{arXiv preprint arXiv:1803.08375}}
  (\bibinfo{year}{2018}).
\newblock


\bibitem[\protect\citeauthoryear{Ahn, Lee, and MacDonald}{Ahn
  et~al\mbox{.}}{2015}]%
        {ahn2015healthcare}
\bibfield{author}{\bibinfo{person}{Ho~Seok Ahn}, \bibinfo{person}{Min~Ho Lee},
  {and} \bibinfo{person}{Bruce~A MacDonald}.} \bibinfo{year}{2015}\natexlab{}.
\newblock \showarticletitle{Healthcare robot systems for a hospital
  environment: CareBot and ReceptionBot}. In \bibinfo{booktitle}{\emph{2015
  24th IEEE International Symposium on Robot and Human Interactive
  Communication (RO-MAN)}}. IEEE, \bibinfo{pages}{571--576}.
\newblock


\bibitem[\protect\citeauthoryear{Al-Jabir, Kerwan, Nicola, Alsafi, Khan,
  Sohrabi, O'Neill, Iosifidis, Griffin, Mathew, and Agha}{Al-Jabir
  et~al\mbox{.}}{2020}]%
        {pmid32407799}
\bibfield{author}{\bibinfo{person}{A. Al-Jabir}, \bibinfo{person}{A. Kerwan},
  \bibinfo{person}{M. Nicola}, \bibinfo{person}{Z. Alsafi}, \bibinfo{person}{M.
  Khan}, \bibinfo{person}{C. Sohrabi}, \bibinfo{person}{N. O'Neill},
  \bibinfo{person}{C. Iosifidis}, \bibinfo{person}{M. Griffin},
  \bibinfo{person}{G. Mathew}, {and} \bibinfo{person}{R. Agha}.}
  \bibinfo{year}{2020}\natexlab{}.
\newblock \showarticletitle{{{I}mpact of the {C}oronavirus ({C}{O}{V}{I}{D}-19)
  pandemic on surgical practice - {P}art 1}}.
\newblock \bibinfo{journal}{\emph{Int J Surg}}  \bibinfo{volume}{79}
  (\bibinfo{date}{Jul} \bibinfo{year}{2020}), \bibinfo{pages}{168--179}.
\newblock


\bibitem[\protect\citeauthoryear{Amirzade~Dana, Esmaeilbeig, and
  Sadeghi}{Amirzade~Dana et~al\mbox{.}}{2020}]%
        {amirzade2020reliability}
\bibfield{author}{\bibinfo{person}{Parvaneh Amirzade~Dana},
  \bibinfo{person}{Zahra Esmaeilbeig}, {and} \bibinfo{person}{Mohammad-Reza
  Sadeghi}.} \bibinfo{year}{2020}\natexlab{}.
\newblock \showarticletitle{Reliability enhancement and packet loss recovery of
  any steganographic method in voice over IP}.
\newblock \bibinfo{journal}{\emph{Wireless Networks}} \bibinfo{volume}{26},
  \bibinfo{number}{8} (\bibinfo{year}{2020}), \bibinfo{pages}{5817--5823}.
\newblock


\bibitem[\protect\citeauthoryear{Anand and Saxena}{Anand and Saxena}{2016}]%
        {anand2016sound}
\bibfield{author}{\bibinfo{person}{S~Abhishek Anand} {and}
  \bibinfo{person}{Nitesh Saxena}.} \bibinfo{year}{2016}\natexlab{}.
\newblock \showarticletitle{A sound for a sound: Mitigating acoustic side
  channel attacks on password keystrokes with active sounds}. In
  \bibinfo{booktitle}{\emph{International Conference on Financial Cryptography
  and Data Security}}. Springer, \bibinfo{pages}{346--364}.
\newblock


\bibitem[\protect\citeauthoryear{Aschenbrenner, Fritscher, Sittner, Krau{\ss},
  and Schilling}{Aschenbrenner et~al\mbox{.}}{2015}]%
        {aschenbrenner2015teleoperation}
\bibfield{author}{\bibinfo{person}{Doris Aschenbrenner},
  \bibinfo{person}{Michael Fritscher}, \bibinfo{person}{Felix Sittner},
  \bibinfo{person}{Markus Krau{\ss}}, {and} \bibinfo{person}{Klaus Schilling}.}
  \bibinfo{year}{2015}\natexlab{}.
\newblock \showarticletitle{Teleoperation of an industrial robot in an active
  production line}.
\newblock \bibinfo{journal}{\emph{IFAC-PapersOnLine}} \bibinfo{volume}{48},
  \bibinfo{number}{10} (\bibinfo{year}{2015}), \bibinfo{pages}{159--164}.
\newblock


\bibitem[\protect\citeauthoryear{Atahan, Elbir, Keskin, Kiraz, Kirval, and
  Aydin}{Atahan et~al\mbox{.}}{2021}]%
        {atahan2021music}
\bibfield{author}{\bibinfo{person}{Yunus Atahan}, \bibinfo{person}{Ahmet
  Elbir}, \bibinfo{person}{Abdullah~Enes Keskin}, \bibinfo{person}{Osman
  Kiraz}, \bibinfo{person}{Bulent Kirval}, {and} \bibinfo{person}{Nizamettin
  Aydin}.} \bibinfo{year}{2021}\natexlab{}.
\newblock \showarticletitle{Music Genre Classification Using Acoustic Features
  and Autoencoders}. In \bibinfo{booktitle}{\emph{2021 Innovations in
  Intelligent Systems and Applications Conference (ASYU)}}. IEEE,
  \bibinfo{pages}{1--5}.
\newblock


\bibitem[\protect\citeauthoryear{Avila, Jimenez, Marquez, Mu{\~n}oz, Carrazco,
  Perdomo, Miselem, and Nolasco}{Avila et~al\mbox{.}}{2020}]%
        {avila2020study}
\bibfield{author}{\bibinfo{person}{Jose Luis~Ordo{\~n}ez Avila},
  \bibinfo{person}{Hector Jimenez}, \bibinfo{person}{Tania Marquez},
  \bibinfo{person}{Carlos Mu{\~n}oz}, \bibinfo{person}{Alberto~Max Carrazco},
  \bibinfo{person}{Maria~Elena Perdomo}, \bibinfo{person}{David Miselem}, {and}
  \bibinfo{person}{David Nolasco}.} \bibinfo{year}{2020}\natexlab{}.
\newblock \showarticletitle{Study Case: Teleoperated Voice Picking Robots
  prototype as a logistic solution in Honduras}. In
  \bibinfo{booktitle}{\emph{2020 5th International Conference on Control and
  Robotics Engineering (ICCRE)}}. IEEE, \bibinfo{pages}{19--24}.
\newblock


\bibitem[\protect\citeauthoryear{Bachu, Kopparthi, Adapa, and Barkana}{Bachu
  et~al\mbox{.}}{2008}]%
        {bachu2008separation}
\bibfield{author}{\bibinfo{person}{RG Bachu}, \bibinfo{person}{S Kopparthi},
  \bibinfo{person}{B Adapa}, {and} \bibinfo{person}{BD Barkana}.}
  \bibinfo{year}{2008}\natexlab{}.
\newblock \showarticletitle{Separation of voiced and unvoiced using zero
  crossing rate and energy of the speech signal}. In
  \bibinfo{booktitle}{\emph{American Society for Engineering Education (ASEE)
  zone conference proceedings}}. American Society for Engineering Education,
  \bibinfo{pages}{1--7}.
\newblock


\bibitem[\protect\citeauthoryear{Backes, D{\"u}rmuth, Gerling, Pinkal,
  Sporleder, et~al\mbox{.}}{Backes et~al\mbox{.}}{2010}]%
        {backes2010acoustic}
\bibfield{author}{\bibinfo{person}{Michael Backes}, \bibinfo{person}{Markus
  D{\"u}rmuth}, \bibinfo{person}{Sebastian Gerling}, \bibinfo{person}{Manfred
  Pinkal}, \bibinfo{person}{Caroline Sporleder}, {et~al\mbox{.}}}
  \bibinfo{year}{2010}\natexlab{}.
\newblock \showarticletitle{Acoustic $\{$Side-Channel$\}$ Attacks on Printers}.
  In \bibinfo{booktitle}{\emph{19th USENIX Security Symposium (USENIX Security
  10)}}.
\newblock


\bibitem[\protect\citeauthoryear{Barto{\v{s}}, Bulej, Bohu{\v{s}}{\'\i}k,
  Stan{\v{c}}ek, Ivanov, and Macek}{Barto{\v{s}} et~al\mbox{.}}{2021}]%
        {bartovs2021overview}
\bibfield{author}{\bibinfo{person}{Michal Barto{\v{s}}},
  \bibinfo{person}{Vladim{\'\i}r Bulej}, \bibinfo{person}{Martin
  Bohu{\v{s}}{\'\i}k}, \bibinfo{person}{J{\'a}n Stan{\v{c}}ek},
  \bibinfo{person}{Vitalii Ivanov}, {and} \bibinfo{person}{Peter Macek}.}
  \bibinfo{year}{2021}\natexlab{}.
\newblock \showarticletitle{An overview of robot applications in automotive
  industry}.
\newblock \bibinfo{journal}{\emph{Transportation Research Procedia}}
  \bibinfo{volume}{55} (\bibinfo{year}{2021}), \bibinfo{pages}{837--844}.
\newblock


\bibitem[\protect\citeauthoryear{Bilal}{Bilal}{2017}]%
        {bilal2017review}
\bibfield{author}{\bibinfo{person}{Muhammad Bilal}.}
  \bibinfo{year}{2017}\natexlab{}.
\newblock \showarticletitle{A review of internet of things architecture,
  technologies and analysis smartphone-based attacks against 3D printers}.
\newblock \bibinfo{journal}{\emph{arXiv preprint arXiv:1708.04560}}
  (\bibinfo{year}{2017}).
\newblock


\bibitem[\protect\citeauthoryear{Bonaci, Herron, Yusuf, Yan, Kohno, and
  Chizeck}{Bonaci et~al\mbox{.}}{2015}]%
        {bonaci2015make}
\bibfield{author}{\bibinfo{person}{Tamara Bonaci}, \bibinfo{person}{Jeffrey
  Herron}, \bibinfo{person}{Tariq Yusuf}, \bibinfo{person}{Junjie Yan},
  \bibinfo{person}{Tadayoshi Kohno}, {and} \bibinfo{person}{Howard~Jay
  Chizeck}.} \bibinfo{year}{2015}\natexlab{}.
\newblock \showarticletitle{To make a robot secure: An experimental analysis of
  cyber security threats against teleoperated surgical robots}.
\newblock \bibinfo{journal}{\emph{arXiv preprint arXiv:1504.04339}}
  (\bibinfo{year}{2015}).
\newblock


\bibitem[\protect\citeauthoryear{Castro}{Castro}{2020}]%
        {castro2020your}
\bibfield{author}{\bibinfo{person}{Rodolfo Castro}.}
  \bibinfo{year}{2020}\natexlab{}.
\newblock \bibinfo{title}{Is your company’s network ready for Microsoft
  teams}.
\newblock
\newblock


\bibitem[\protect\citeauthoryear{Chhetri, Canedo, and Faruque}{Chhetri
  et~al\mbox{.}}{2017}]%
        {chhetri2017confidentiality}
\bibfield{author}{\bibinfo{person}{Sujit~Rokka Chhetri},
  \bibinfo{person}{Arquimedes Canedo}, {and} \bibinfo{person}{Mohammad
  Abdullah~Al Faruque}.} \bibinfo{year}{2017}\natexlab{}.
\newblock \showarticletitle{Confidentiality breach through acoustic
  side-channel in cyber-physical additive manufacturing systems}.
\newblock \bibinfo{journal}{\emph{ACM Transactions on Cyber-Physical Systems}}
  \bibinfo{volume}{2}, \bibinfo{number}{1} (\bibinfo{year}{2017}),
  \bibinfo{pages}{1--25}.
\newblock


\bibitem[\protect\citeauthoryear{Cho and Bello}{Cho and Bello}{2013}]%
        {cho2013relative}
\bibfield{author}{\bibinfo{person}{Taemin Cho} {and} \bibinfo{person}{Juan~P
  Bello}.} \bibinfo{year}{2013}\natexlab{}.
\newblock \showarticletitle{On the relative importance of individual components
  of chord recognition systems}.
\newblock \bibinfo{journal}{\emph{IEEE/ACM Transactions on Audio, Speech, and
  Language Processing}} \bibinfo{volume}{22}, \bibinfo{number}{2}
  (\bibinfo{year}{2013}), \bibinfo{pages}{477--492}.
\newblock


\bibitem[\protect\citeauthoryear{Chollet et~al\mbox{.}}{Chollet
  et~al\mbox{.}}{2015}]%
        {chollet2015keras}
\bibfield{author}{\bibinfo{person}{Fran\c{c}ois Chollet} {et~al\mbox{.}}}
  \bibinfo{year}{2015}\natexlab{}.
\newblock \bibinfo{title}{Keras}.
\newblock \bibinfo{howpublished}{\url{https://keras.io}}.
\newblock


\bibitem[\protect\citeauthoryear{Dalen, Legemaate, Schlack, Legemate, and
  Schijven}{Dalen et~al\mbox{.}}{2019}]%
        {dalen2019legal}
\bibfield{author}{\bibinfo{person}{ASHM Dalen}, \bibinfo{person}{J Legemaate},
  \bibinfo{person}{WS Schlack}, \bibinfo{person}{DA Legemate}, {and}
  \bibinfo{person}{MP Schijven}.} \bibinfo{year}{2019}\natexlab{}.
\newblock \showarticletitle{Legal perspectives on black box recording devices
  in the operating environment}.
\newblock \bibinfo{journal}{\emph{Journal of British Surgery}}
  \bibinfo{volume}{106}, \bibinfo{number}{11} (\bibinfo{year}{2019}),
  \bibinfo{pages}{1433--1441}.
\newblock


\bibitem[\protect\citeauthoryear{Demuth, Beale, De~Jess, and Hagan}{Demuth
  et~al\mbox{.}}{2014}]%
        {demuth2014neural}
\bibfield{author}{\bibinfo{person}{Howard~B Demuth}, \bibinfo{person}{Mark~H
  Beale}, \bibinfo{person}{Orlando De~Jess}, {and} \bibinfo{person}{Martin~T
  Hagan}.} \bibinfo{year}{2014}\natexlab{}.
\newblock \bibinfo{booktitle}{\emph{Neural network design}}.
\newblock \bibinfo{publisher}{Martin Hagan}.
\newblock


\bibitem[\protect\citeauthoryear{Desai and Patil}{Desai and Patil}{[n.d.]}]%
        {desaiexperimental}
\bibfield{author}{\bibinfo{person}{Tanvi~D Desai} {and} \bibinfo{person}{SR
  Patil}.} \bibinfo{year}{[n.d.]}\natexlab{}.
\newblock \showarticletitle{Experimental and Numerical Analysis of Vibration
  Isolation Materials on Vibration Reduction within Plazma Torch}.
\newblock  (\bibinfo{year}{[n.\,d.]}).
\newblock


\bibitem[\protect\citeauthoryear{Doppler}{Doppler}{1903}]%
        {doppler1903ueber}
\bibfield{author}{\bibinfo{person}{Christian Doppler}.}
  \bibinfo{year}{1903}\natexlab{}.
\newblock \bibinfo{booktitle}{\emph{Ueber das farbige Licht der Doppelsterne
  und einiger anderer Gestirne des Himmels: Versuch einer das Bradley'sche
  Aberrations-Theorem als integrirenden Theil in sich schliessenden
  allgemeineren Theorie}}.
\newblock \bibinfo{publisher}{K. B{\"o}hm Gesellschaft der Wissenschaften}.
\newblock


\bibitem[\protect\citeauthoryear{Dunne and Campbell}{Dunne and
  Campbell}{1997}]%
        {dunne1997pairing}
\bibfield{author}{\bibinfo{person}{Rob~A Dunne} {and} \bibinfo{person}{Norm~A
  Campbell}.} \bibinfo{year}{1997}\natexlab{}.
\newblock \showarticletitle{On the pairing of the softmax activation and
  cross-entropy penalty functions and the derivation of the softmax activation
  function}. In \bibinfo{booktitle}{\emph{Proc. 8th Aust. Conf. on the Neural
  Networks, Melbourne}}, Vol.~\bibinfo{volume}{181}. Citeseer,
  \bibinfo{pages}{185}.
\newblock


\bibitem[\protect\citeauthoryear{Eckle and Schmidt-Hieber}{Eckle and
  Schmidt-Hieber}{2019}]%
        {eckle2019comparison}
\bibfield{author}{\bibinfo{person}{Konstantin Eckle} {and}
  \bibinfo{person}{Johannes Schmidt-Hieber}.} \bibinfo{year}{2019}\natexlab{}.
\newblock \showarticletitle{A comparison of deep networks with ReLU activation
  function and linear spline-type methods}.
\newblock \bibinfo{journal}{\emph{Neural Networks}}  \bibinfo{volume}{110}
  (\bibinfo{year}{2019}), \bibinfo{pages}{232--242}.
\newblock


\bibitem[\protect\citeauthoryear{Grabowski, Jankowski, and
  Wodzy{\'n}ski}{Grabowski et~al\mbox{.}}{2021}]%
        {grabowski2021teleoperated}
\bibfield{author}{\bibinfo{person}{Andrzej Grabowski},
  \bibinfo{person}{Jaros{\l}aw Jankowski}, {and} \bibinfo{person}{Mieszko
  Wodzy{\'n}ski}.} \bibinfo{year}{2021}\natexlab{}.
\newblock \showarticletitle{Teleoperated mobile robot with two arms: the
  influence of a human-machine interface, VR training and operator age}.
\newblock \bibinfo{journal}{\emph{International Journal of Human-Computer
  Studies}}  \bibinfo{volume}{156} (\bibinfo{year}{2021}),
  \bibinfo{pages}{102707}.
\newblock


\bibitem[\protect\citeauthoryear{Gravagne, Rahn, and Walker}{Gravagne
  et~al\mbox{.}}{2001}]%
        {gravagne2001good}
\bibfield{author}{\bibinfo{person}{Ian~A Gravagne},
  \bibinfo{person}{Christopher~D Rahn}, {and} \bibinfo{person}{Ian~D Walker}.}
  \bibinfo{year}{2001}\natexlab{}.
\newblock \showarticletitle{Good vibrations: a vibration damping setpoint
  controller for continuum robots}. In \bibinfo{booktitle}{\emph{Proceedings
  2001 ICRA. IEEE International Conference on Robotics and Automation (Cat. No.
  01CH37164)}}, Vol.~\bibinfo{volume}{4}. IEEE, \bibinfo{pages}{3877--3884}.
\newblock


\bibitem[\protect\citeauthoryear{Greenwood}{Greenwood}{1997}]%
        {greenwood1997mel}
\bibfield{author}{\bibinfo{person}{Donald~D Greenwood}.}
  \bibinfo{year}{1997}\natexlab{}.
\newblock \showarticletitle{The Mel Scale's disqualifying bias and a
  consistency of pitch-difference equisections in 1956 with equal cochlear
  distances and equal frequency ratios}.
\newblock \bibinfo{journal}{\emph{Hearing research}} \bibinfo{volume}{103},
  \bibinfo{number}{1-2} (\bibinfo{year}{1997}), \bibinfo{pages}{199--224}.
\newblock


\bibitem[\protect\citeauthoryear{Hannaford, Rosen, Friedman, King, Roan, Cheng,
  Glozman, Ma, Kosari, and White}{Hannaford et~al\mbox{.}}{2012}]%
        {hannaford2012raven}
\bibfield{author}{\bibinfo{person}{Blake Hannaford}, \bibinfo{person}{Jacob
  Rosen}, \bibinfo{person}{Diana~W Friedman}, \bibinfo{person}{Hawkeye King},
  \bibinfo{person}{Phillip Roan}, \bibinfo{person}{Lei Cheng},
  \bibinfo{person}{Daniel Glozman}, \bibinfo{person}{Ji Ma},
  \bibinfo{person}{Sina~Nia Kosari}, {and} \bibinfo{person}{Lee White}.}
  \bibinfo{year}{2012}\natexlab{}.
\newblock \showarticletitle{Raven-II: an open platform for surgical robotics
  research}.
\newblock \bibinfo{journal}{\emph{IEEE Transactions on Biomedical Engineering}}
  \bibinfo{volume}{60}, \bibinfo{number}{4} (\bibinfo{year}{2012}),
  \bibinfo{pages}{954--959}.
\newblock


\bibitem[\protect\citeauthoryear{Hosseini, Moghaddasi, Sajadi, and
  Karimi}{Hosseini et~al\mbox{.}}{2013}]%
        {hosseini2013telesurgery}
\bibfield{author}{\bibinfo{person}{Azamossadat Hosseini},
  \bibinfo{person}{Hamid Moghaddasi}, \bibinfo{person}{Samad Sajadi}, {and}
  \bibinfo{person}{Mozhgan Karimi}.} \bibinfo{year}{2013}\natexlab{}.
\newblock \showarticletitle{Telesurgery information management systems in
  university hospitals of Tehran}.
\newblock \bibinfo{journal}{\emph{Archives of Advances in Biosciences}}
  \bibinfo{volume}{4}, \bibinfo{number}{4} (\bibinfo{year}{2013}).
\newblock


\bibitem[\protect\citeauthoryear{Inouye, Blemker, and Inouye}{Inouye
  et~al\mbox{.}}{2014}]%
        {inouye2014towards}
\bibfield{author}{\bibinfo{person}{Joshua~M Inouye}, \bibinfo{person}{Silvia~S
  Blemker}, {and} \bibinfo{person}{David~I Inouye}.}
  \bibinfo{year}{2014}\natexlab{}.
\newblock \showarticletitle{Towards undistorted and noise-free speech in an MRI
  scanner: correlation subtraction followed by spectral noise gating}.
\newblock \bibinfo{journal}{\emph{The Journal of the Acoustical Society of
  America}} \bibinfo{volume}{135}, \bibinfo{number}{3} (\bibinfo{year}{2014}),
  \bibinfo{pages}{1019--1022}.
\newblock


\bibitem[\protect\citeauthoryear{James, Chen, and Garrison}{James
  et~al\mbox{.}}{2004}]%
        {james2004implementing}
\bibfield{author}{\bibinfo{person}{Jim~H James}, \bibinfo{person}{Bing Chen},
  {and} \bibinfo{person}{Laurie Garrison}.} \bibinfo{year}{2004}\natexlab{}.
\newblock \showarticletitle{Implementing VoIP: a voice transmission performance
  progress report}.
\newblock \bibinfo{journal}{\emph{IEEE Communications Magazine}}
  \bibinfo{volume}{42}, \bibinfo{number}{7} (\bibinfo{year}{2004}),
  \bibinfo{pages}{36--41}.
\newblock


\bibitem[\protect\citeauthoryear{Jiang, Lu, Zhang, Tao, and Cai}{Jiang
  et~al\mbox{.}}{2002}]%
        {jiang2002music}
\bibfield{author}{\bibinfo{person}{Dan-Ning Jiang}, \bibinfo{person}{Lie Lu},
  \bibinfo{person}{Hong-Jiang Zhang}, \bibinfo{person}{Jian-Hua Tao}, {and}
  \bibinfo{person}{Lian-Hong Cai}.} \bibinfo{year}{2002}\natexlab{}.
\newblock \showarticletitle{Music type classification by spectral contrast
  feature}. In \bibinfo{booktitle}{\emph{Proceedings. IEEE International
  Conference on Multimedia and Expo}}, Vol.~\bibinfo{volume}{1}. IEEE,
  \bibinfo{pages}{113--116}.
\newblock


\bibitem[\protect\citeauthoryear{Khan and Li}{Khan and Li}{2020}]%
        {khan2020sliding}
\bibfield{author}{\bibinfo{person}{Ameer~Hamza Khan} {and}
  \bibinfo{person}{Shuai Li}.} \bibinfo{year}{2020}\natexlab{}.
\newblock \showarticletitle{Sliding mode control with PID sliding surface for
  active vibration damping of pneumatically actuated soft robots}.
\newblock \bibinfo{journal}{\emph{IEEE Access}}  \bibinfo{volume}{8}
  (\bibinfo{year}{2020}), \bibinfo{pages}{88793--88800}.
\newblock


\bibitem[\protect\citeauthoryear{Kim, Lee, Raghunathan, Jha, and
  Raghunathan}{Kim et~al\mbox{.}}{2015}]%
        {kim2015vibration}
\bibfield{author}{\bibinfo{person}{Younghyun Kim}, \bibinfo{person}{Woo~Suk
  Lee}, \bibinfo{person}{Vijay Raghunathan}, \bibinfo{person}{Niraj~K Jha},
  {and} \bibinfo{person}{Anand Raghunathan}.} \bibinfo{year}{2015}\natexlab{}.
\newblock \showarticletitle{Vibration-based secure side channel for medical
  devices}. In \bibinfo{booktitle}{\emph{2015 52nd ACM/EDAC/IEEE Design
  Automation Conference (DAC)}}. IEEE, \bibinfo{pages}{1--6}.
\newblock


\bibitem[\protect\citeauthoryear{Kingma and Ba}{Kingma and Ba}{2014}]%
        {kingma2014adam}
\bibfield{author}{\bibinfo{person}{Diederik~P Kingma} {and}
  \bibinfo{person}{Jimmy Ba}.} \bibinfo{year}{2014}\natexlab{}.
\newblock \showarticletitle{Adam: A method for stochastic optimization}.
\newblock \bibinfo{journal}{\emph{arXiv preprint arXiv:1412.6980}}
  (\bibinfo{year}{2014}).
\newblock


\bibitem[\protect\citeauthoryear{Klapuri and Davy}{Klapuri and Davy}{2007}]%
        {klapuri2007signal}
\bibfield{author}{\bibinfo{person}{Anssi Klapuri} {and} \bibinfo{person}{Manuel
  Davy}.} \bibinfo{year}{2007}\natexlab{}.
\newblock \showarticletitle{Signal processing methods for music transcription}.
\newblock  (\bibinfo{year}{2007}).
\newblock


\bibitem[\protect\citeauthoryear{Kos, Ka{\v{c}}i{\v{c}}, and Vlaj}{Kos
  et~al\mbox{.}}{2013}]%
        {kos2013acoustic}
\bibfield{author}{\bibinfo{person}{Marko Kos}, \bibinfo{person}{Zdravko
  Ka{\v{c}}i{\v{c}}}, {and} \bibinfo{person}{Damjan Vlaj}.}
  \bibinfo{year}{2013}\natexlab{}.
\newblock \showarticletitle{Acoustic classification and segmentation using
  modified spectral roll-off and variance-based features}.
\newblock \bibinfo{journal}{\emph{Digital Signal Processing}}
  \bibinfo{volume}{23}, \bibinfo{number}{2} (\bibinfo{year}{2013}),
  \bibinfo{pages}{659--674}.
\newblock


\bibitem[\protect\citeauthoryear{Krizhevsky, Sutskever, and Hinton}{Krizhevsky
  et~al\mbox{.}}{2012}]%
        {krizhevsky2012imagenet}
\bibfield{author}{\bibinfo{person}{Alex Krizhevsky}, \bibinfo{person}{Ilya
  Sutskever}, {and} \bibinfo{person}{Geoffrey~E Hinton}.}
  \bibinfo{year}{2012}\natexlab{}.
\newblock \showarticletitle{Imagenet classification with deep convolutional
  neural networks}. In \bibinfo{booktitle}{\emph{Advances in neural information
  processing systems}}. \bibinfo{pages}{1097--1105}.
\newblock


\bibitem[\protect\citeauthoryear{Kulkarni, Torse, and Kulkarni}{Kulkarni
  et~al\mbox{.}}{2020}]%
        {kulkarni2020cloud}
\bibfield{author}{\bibinfo{person}{Sushmita Kulkarni},
  \bibinfo{person}{Dattaprasad~A Torse}, {and} \bibinfo{person}{Deepak
  Kulkarni}.} \bibinfo{year}{2020}\natexlab{}.
\newblock \showarticletitle{A Cloud based Medical Transcription using Speech
  Recognition Technologies}.
\newblock \bibinfo{journal}{\emph{International Research Journal of Engineering
  and Technology (IRJET)}} \bibinfo{volume}{7}, \bibinfo{number}{5}
  (\bibinfo{year}{2020}), \bibinfo{pages}{6160--6163}.
\newblock


\bibitem[\protect\citeauthoryear{Laghari, Laghari, Wagan, and Umrani}{Laghari
  et~al\mbox{.}}{2020}]%
        {laghari2020effect}
\bibfield{author}{\bibinfo{person}{Asif~Ali Laghari},
  \bibinfo{person}{Rashid~Ali Laghari}, \bibinfo{person}{Asif~Ali Wagan}, {and}
  \bibinfo{person}{Aamir~Iqbal Umrani}.} \bibinfo{year}{2020}\natexlab{}.
\newblock \showarticletitle{Effect of packet loss and reorder on quality of
  audio streaming}.
\newblock \bibinfo{journal}{\emph{EAI Endorsed Transactions on Scalable
  Information Systems}} \bibinfo{volume}{7}, \bibinfo{number}{25}
  (\bibinfo{year}{2020}).
\newblock


\bibitem[\protect\citeauthoryear{Le, Ambikairajah, Epps, Sethu, and Choi}{Le
  et~al\mbox{.}}{2011}]%
        {le2011investigation}
\bibfield{author}{\bibinfo{person}{Phu~Ngoc Le}, \bibinfo{person}{Eliathamby
  Ambikairajah}, \bibinfo{person}{Julien Epps}, \bibinfo{person}{Vidhyasaharan
  Sethu}, {and} \bibinfo{person}{Eric~HC Choi}.}
  \bibinfo{year}{2011}\natexlab{}.
\newblock \showarticletitle{Investigation of spectral centroid features for
  cognitive load classification}.
\newblock \bibinfo{journal}{\emph{Speech Communication}} \bibinfo{volume}{53},
  \bibinfo{number}{4} (\bibinfo{year}{2011}), \bibinfo{pages}{540--551}.
\newblock


\bibitem[\protect\citeauthoryear{Li, Yang, Wan, Annamalai, and Cangelosi}{Li
  et~al\mbox{.}}{2017}]%
        {li2017teleoperation}
\bibfield{author}{\bibinfo{person}{Chunxu Li}, \bibinfo{person}{Chenguang
  Yang}, \bibinfo{person}{Jian Wan}, \bibinfo{person}{Andy~SK Annamalai}, {and}
  \bibinfo{person}{Angelo Cangelosi}.} \bibinfo{year}{2017}\natexlab{}.
\newblock \showarticletitle{Teleoperation control of Baxter robot using Kalman
  filter-based sensor fusion}.
\newblock \bibinfo{journal}{\emph{Systems Science \& Control Engineering}}
  \bibinfo{volume}{5}, \bibinfo{number}{1} (\bibinfo{year}{2017}),
  \bibinfo{pages}{156--167}.
\newblock


\bibitem[\protect\citeauthoryear{Li and Yuan}{Li and Yuan}{2017}]%
        {li2017convergence}
\bibfield{author}{\bibinfo{person}{Yuanzhi Li} {and} \bibinfo{person}{Yang
  Yuan}.} \bibinfo{year}{2017}\natexlab{}.
\newblock \showarticletitle{Convergence analysis of two-layer neural networks
  with relu activation}. In \bibinfo{booktitle}{\emph{Advances in neural
  information processing systems}}. \bibinfo{pages}{597--607}.
\newblock


\bibitem[\protect\citeauthoryear{Ma, Wong, and Zhao}{Ma et~al\mbox{.}}{2019}]%
        {ma2019practical}
\bibfield{author}{\bibinfo{person}{Xinbo Ma}, \bibinfo{person}{Pak~Kin Wong},
  {and} \bibinfo{person}{Jing Zhao}.} \bibinfo{year}{2019}\natexlab{}.
\newblock \showarticletitle{Practical multi-objective control for automotive
  semi-active suspension system with nonlinear hydraulic adjustable damper}.
\newblock \bibinfo{journal}{\emph{Mechanical Systems and Signal Processing}}
  \bibinfo{volume}{117} (\bibinfo{year}{2019}), \bibinfo{pages}{667--688}.
\newblock


\bibitem[\protect\citeauthoryear{Martinez, Perez, Escamilla, and
  Suzuki}{Martinez et~al\mbox{.}}{2012}]%
        {martinez2012speaker}
\bibfield{author}{\bibinfo{person}{Jorge Martinez}, \bibinfo{person}{Hector
  Perez}, \bibinfo{person}{Enrique Escamilla}, {and}
  \bibinfo{person}{Masahisa~Mabo Suzuki}.} \bibinfo{year}{2012}\natexlab{}.
\newblock \showarticletitle{Speaker recognition using Mel frequency Cepstral
  Coefficients (MFCC) and Vector quantization (VQ) techniques}. In
  \bibinfo{booktitle}{\emph{CONIELECOMP 2012, 22nd International Conference on
  Electrical Communications and Computers}}. IEEE, \bibinfo{pages}{248--251}.
\newblock


\bibitem[\protect\citeauthoryear{McClean, Stull, Farrar, and
  Mascarenas}{McClean et~al\mbox{.}}{2013}]%
        {mcclean2013preliminary}
\bibfield{author}{\bibinfo{person}{Jarrod McClean},
  \bibinfo{person}{Christopher Stull}, \bibinfo{person}{Charles Farrar}, {and}
  \bibinfo{person}{David Mascarenas}.} \bibinfo{year}{2013}\natexlab{}.
\newblock \showarticletitle{A preliminary cyber-physical security assessment of
  the robot operating system (ros)}. In \bibinfo{booktitle}{\emph{Unmanned
  Systems Technology XV}}, Vol.~\bibinfo{volume}{8741}. International Society
  for Optics and Photonics, \bibinfo{pages}{874110}.
\newblock


\bibitem[\protect\citeauthoryear{McFee, Raffel, Liang, Ellis, McVicar,
  Battenberg, and Nieto}{McFee et~al\mbox{.}}{2015}]%
        {mcfee2015librosa}
\bibfield{author}{\bibinfo{person}{Brian McFee}, \bibinfo{person}{Colin
  Raffel}, \bibinfo{person}{Dawen Liang}, \bibinfo{person}{Daniel~P Ellis},
  \bibinfo{person}{Matt McVicar}, \bibinfo{person}{Eric Battenberg}, {and}
  \bibinfo{person}{Oriol Nieto}.} \bibinfo{year}{2015}\natexlab{}.
\newblock \showarticletitle{librosa: Audio and music signal analysis in
  python}. In \bibinfo{booktitle}{\emph{Proceedings of the 14th python in
  science conference}}, Vol.~\bibinfo{volume}{8}. Citeseer,
  \bibinfo{pages}{18--25}.
\newblock


\bibitem[\protect\citeauthoryear{Muensterer, Lacher, Zoeller, Bronstein, and
  K{\"u}bler}{Muensterer et~al\mbox{.}}{2014}]%
        {muensterer2014google}
\bibfield{author}{\bibinfo{person}{Oliver~J Muensterer},
  \bibinfo{person}{Martin Lacher}, \bibinfo{person}{Christoph Zoeller},
  \bibinfo{person}{Matthew Bronstein}, {and} \bibinfo{person}{Joachim
  K{\"u}bler}.} \bibinfo{year}{2014}\natexlab{}.
\newblock \showarticletitle{Google Glass in pediatric surgery: an exploratory
  study}.
\newblock \bibinfo{journal}{\emph{International journal of surgery}}
  \bibinfo{volume}{12}, \bibinfo{number}{4} (\bibinfo{year}{2014}),
  \bibinfo{pages}{281--289}.
\newblock


\bibitem[\protect\citeauthoryear{M{\"u}ller}{M{\"u}ller}{2015}]%
        {muller2015fundamentals}
\bibfield{author}{\bibinfo{person}{Meinard M{\"u}ller}.}
  \bibinfo{year}{2015}\natexlab{}.
\newblock \showarticletitle{Fundamentals of Music Processing}.
\newblock  (\bibinfo{year}{2015}).
\newblock


\bibitem[\protect\citeauthoryear{Neumann and Schuller}{Neumann and
  Schuller}{1991}]%
        {neumann1991spectral}
\bibfield{author}{\bibinfo{person}{Ingrid Neumann} {and} \bibinfo{person}{Gerd
  Schuller}.} \bibinfo{year}{1991}\natexlab{}.
\newblock \showarticletitle{Spectral and temporal gating mechanisms enhance the
  clutter rejection in the echolocating bat, Rhinolophus rouxi}.
\newblock \bibinfo{journal}{\emph{Journal of comparative physiology A}}
  \bibinfo{volume}{169}, \bibinfo{number}{1} (\bibinfo{year}{1991}),
  \bibinfo{pages}{109--116}.
\newblock


\bibitem[\protect\citeauthoryear{Ortega, Altamirano, and Abad}{Ortega
  et~al\mbox{.}}{2018}]%
        {ortega2018evaluation}
\bibfield{author}{\bibinfo{person}{Mart{\'\i}n~Ortega Ortega},
  \bibinfo{person}{Gustavo~Chafla Altamirano}, {and}
  \bibinfo{person}{Mara~Falcon{\'\i} Abad}.} \bibinfo{year}{2018}\natexlab{}.
\newblock \showarticletitle{Evaluation of the voice quality and QoS in real
  calls using different voice over IP codecs}. In
  \bibinfo{booktitle}{\emph{2018 IEEE Colombian Conference on Communications
  and Computing (COLCOM)}}. IEEE, \bibinfo{pages}{1--6}.
\newblock


\bibitem[\protect\citeauthoryear{Panagiotakis and Tziritas}{Panagiotakis and
  Tziritas}{2005}]%
        {panagiotakis2005speech}
\bibfield{author}{\bibinfo{person}{Costas Panagiotakis} {and}
  \bibinfo{person}{Georgios Tziritas}.} \bibinfo{year}{2005}\natexlab{}.
\newblock \showarticletitle{A speech/music discriminator based on RMS and
  zero-crossings}.
\newblock \bibinfo{journal}{\emph{IEEE Transactions on multimedia}}
  \bibinfo{volume}{7}, \bibinfo{number}{1} (\bibinfo{year}{2005}),
  \bibinfo{pages}{155--166}.
\newblock


\bibitem[\protect\citeauthoryear{Paulus, M{\"u}ller, and Klapuri}{Paulus
  et~al\mbox{.}}{2010}]%
        {paulus2010state}
\bibfield{author}{\bibinfo{person}{Jouni Paulus}, \bibinfo{person}{Meinard
  M{\"u}ller}, {and} \bibinfo{person}{Anssi Klapuri}.}
  \bibinfo{year}{2010}\natexlab{}.
\newblock \showarticletitle{State of the Art Report: Audio-Based Music
  Structure Analysis.}. In \bibinfo{booktitle}{\emph{Ismir}}. Utrecht,
  \bibinfo{pages}{625--636}.
\newblock


\bibitem[\protect\citeauthoryear{Peng and Choi}{Peng and Choi}{2013}]%
        {peng2013mobile}
\bibfield{author}{\bibinfo{person}{Yinni Peng} {and}
  \bibinfo{person}{Susanne~YP Choi}.} \bibinfo{year}{2013}\natexlab{}.
\newblock \showarticletitle{Mobile phone use among migrant factory workers in
  south China: Technologies of power and resistance}.
\newblock \bibinfo{journal}{\emph{The China Quarterly}}  \bibinfo{volume}{215}
  (\bibinfo{year}{2013}), \bibinfo{pages}{553--571}.
\newblock


\bibitem[\protect\citeauthoryear{Polydoros and Nalpantidis}{Polydoros and
  Nalpantidis}{2016}]%
        {polydoros2016reservoir}
\bibfield{author}{\bibinfo{person}{Athanasios~S Polydoros} {and}
  \bibinfo{person}{Lazaros Nalpantidis}.} \bibinfo{year}{2016}\natexlab{}.
\newblock \showarticletitle{A reservoir computing approach for learning forward
  dynamics of industrial manipulators}. In \bibinfo{booktitle}{\emph{2016
  IEEE/RSJ International Conference on Intelligent Robots and Systems (IROS)}}.
  IEEE, \bibinfo{pages}{612--618}.
\newblock


\bibitem[\protect\citeauthoryear{Quarta, Pogliani, Polino, Maggi, Zanchettin,
  and Zanero}{Quarta et~al\mbox{.}}{2017}]%
        {quarta2017experimental}
\bibfield{author}{\bibinfo{person}{Davide Quarta}, \bibinfo{person}{Marcello
  Pogliani}, \bibinfo{person}{Mario Polino}, \bibinfo{person}{Federico Maggi},
  \bibinfo{person}{Andrea~Maria Zanchettin}, {and} \bibinfo{person}{Stefano
  Zanero}.} \bibinfo{year}{2017}\natexlab{}.
\newblock \showarticletitle{An experimental security analysis of an industrial
  robot controller}. In \bibinfo{booktitle}{\emph{2017 IEEE Symposium on
  Security and Privacy (SP)}}. IEEE, \bibinfo{pages}{268--286}.
\newblock


\bibitem[\protect\citeauthoryear{Rajaratnam, Shah, and Kalita}{Rajaratnam
  et~al\mbox{.}}{2018}]%
        {rajaratnam2018isolated}
\bibfield{author}{\bibinfo{person}{Krishan Rajaratnam}, \bibinfo{person}{Kunal
  Shah}, {and} \bibinfo{person}{Jugal Kalita}.}
  \bibinfo{year}{2018}\natexlab{}.
\newblock \showarticletitle{Isolated and ensemble audio preprocessing methods
  for detecting adversarial examples against automatic speech recognition}.
\newblock \bibinfo{journal}{\emph{arXiv preprint arXiv:1809.04397}}
  (\bibinfo{year}{2018}).
\newblock


\bibitem[\protect\citeauthoryear{R{\"a}m{\"o} and Toukomaa}{R{\"a}m{\"o} and
  Toukomaa}{2011}]%
        {ramo2011voice}
\bibfield{author}{\bibinfo{person}{Anssi R{\"a}m{\"o}} {and}
  \bibinfo{person}{Henri Toukomaa}.} \bibinfo{year}{2011}\natexlab{}.
\newblock \showarticletitle{Voice quality characterization of IETF Opus codec}.
  In \bibinfo{booktitle}{\emph{Twelfth Annual Conference of the International
  Speech Communication Association}}.
\newblock


\bibitem[\protect\citeauthoryear{Rao and Vuppala}{Rao and Vuppala}{2014}]%
        {rao2014speech}
\bibfield{author}{\bibinfo{person}{K~Sreenivasa Rao} {and}
  \bibinfo{person}{Anil~Kumar Vuppala}.} \bibinfo{year}{2014}\natexlab{}.
\newblock \bibinfo{booktitle}{\emph{Speech processing in mobile environments}}.
\newblock \bibinfo{publisher}{Springer}.
\newblock


\bibitem[\protect\citeauthoryear{Reuters}{Reuters}{2019}]%
        {reutersrobots}
\bibfield{author}{\bibinfo{person}{Reuters}.} \bibinfo{year}{2019}\natexlab{}.
\newblock \bibinfo{title}{U.S. companies put record number of robots to work in
  2018}.
\newblock
  \bibinfo{howpublished}{\url{https://www.reuters.com/article/us-usa-economy-robots/u-s-companies-put-record-number-of-robots-to-work-in-2018-idUSKCN1QH0K0}}.
\newblock


\bibitem[\protect\citeauthoryear{Rueckert, Nakatenus, Tosatto, and
  Peters}{Rueckert et~al\mbox{.}}{2017}]%
        {rueckert2017learning}
\bibfield{author}{\bibinfo{person}{Elmar Rueckert}, \bibinfo{person}{Moritz
  Nakatenus}, \bibinfo{person}{Samuele Tosatto}, {and} \bibinfo{person}{Jan
  Peters}.} \bibinfo{year}{2017}\natexlab{}.
\newblock \showarticletitle{Learning inverse dynamics models in o (n) time with
  lstm networks}. In \bibinfo{booktitle}{\emph{2017 IEEE-RAS 17th International
  Conference on Humanoid Robotics (Humanoids)}}. IEEE,
  \bibinfo{pages}{811--816}.
\newblock


\bibitem[\protect\citeauthoryear{Saun, Zuo, and Grantcharov}{Saun
  et~al\mbox{.}}{2019}]%
        {saun2019video}
\bibfield{author}{\bibinfo{person}{Tomas~J Saun}, \bibinfo{person}{Kevin~J
  Zuo}, {and} \bibinfo{person}{Teodor~P Grantcharov}.}
  \bibinfo{year}{2019}\natexlab{}.
\newblock \showarticletitle{Video technologies for recording open surgery: a
  systematic review}.
\newblock \bibinfo{journal}{\emph{Surgical innovation}} \bibinfo{volume}{26},
  \bibinfo{number}{5} (\bibinfo{year}{2019}), \bibinfo{pages}{599--612}.
\newblock


\bibitem[\protect\citeauthoryear{Shah, Ahmed, and Nagaraja}{Shah
  et~al\mbox{.}}{2022}]%
        {shah2022can}
\bibfield{author}{\bibinfo{person}{Ryan Shah}, \bibinfo{person}{Chuadhry~Mujeeb
  Ahmed}, {and} \bibinfo{person}{Shishir Nagaraja}.}
  \bibinfo{year}{2022}\natexlab{}.
\newblock \showarticletitle{Can You Still See Me?: Reconstructing Robot
  Operations Over End-to-End Encrypted Channels}.
\newblock \bibinfo{journal}{\emph{arXiv preprint arXiv:2205.08426}}
  (\bibinfo{year}{2022}).
\newblock


\bibitem[\protect\citeauthoryear{Song, Lin, Ba, Ren, Zhou, and Xu}{Song
  et~al\mbox{.}}{2016}]%
        {song2016my}
\bibfield{author}{\bibinfo{person}{Chen Song}, \bibinfo{person}{Feng Lin},
  \bibinfo{person}{Zhongjie Ba}, \bibinfo{person}{Kui Ren},
  \bibinfo{person}{Chi Zhou}, {and} \bibinfo{person}{Wenyao Xu}.}
  \bibinfo{year}{2016}\natexlab{}.
\newblock \showarticletitle{My smartphone knows what you print: Exploring
  smartphone-based side-channel attacks against 3d printers}. In
  \bibinfo{booktitle}{\emph{Proceedings of the 2016 ACM SIGSAC Conference on
  Computer and Communications Security}}. \bibinfo{pages}{895--907}.
\newblock


\bibitem[\protect\citeauthoryear{Sung and Gill}{Sung and Gill}{2001}]%
        {sung2001robotic}
\bibfield{author}{\bibinfo{person}{Gyung~Tak Sung} {and}
  \bibinfo{person}{Inderbir~S Gill}.} \bibinfo{year}{2001}\natexlab{}.
\newblock \showarticletitle{Robotic laparoscopic surgery: a comparison of the
  da Vinci and Zeus systems}.
\newblock \bibinfo{journal}{\emph{Urology}} \bibinfo{volume}{58},
  \bibinfo{number}{6} (\bibinfo{year}{2001}), \bibinfo{pages}{893--898}.
\newblock


\bibitem[\protect\citeauthoryear{Tewari, Peabody, Sarle, Balakrishnan, Hemal,
  Shrivastava, and Menon}{Tewari et~al\mbox{.}}{2002}]%
        {tewari2002technique}
\bibfield{author}{\bibinfo{person}{Ashutosh Tewari}, \bibinfo{person}{James
  Peabody}, \bibinfo{person}{Richard Sarle}, \bibinfo{person}{Guruswami
  Balakrishnan}, \bibinfo{person}{Ashok Hemal}, \bibinfo{person}{Alok
  Shrivastava}, {and} \bibinfo{person}{Mani Menon}.}
  \bibinfo{year}{2002}\natexlab{}.
\newblock \showarticletitle{Technique of da Vinci robot-assisted anatomic
  radical prostatectomy}.
\newblock \bibinfo{journal}{\emph{Urology}} \bibinfo{volume}{60},
  \bibinfo{number}{4} (\bibinfo{year}{2002}), \bibinfo{pages}{569--572}.
\newblock


\bibitem[\protect\citeauthoryear{Valin}{Valin}{2016}]%
        {valin2016speex}
\bibfield{author}{\bibinfo{person}{Jean-Marc Valin}.}
  \bibinfo{year}{2016}\natexlab{}.
\newblock \showarticletitle{Speex: A free codec for free speech}.
\newblock \bibinfo{journal}{\emph{arXiv preprint arXiv:1602.08668}}
  (\bibinfo{year}{2016}).
\newblock


\bibitem[\protect\citeauthoryear{Valin, Maxwell, Terriberry, and Vos}{Valin
  et~al\mbox{.}}{2016}]%
        {valin2016high}
\bibfield{author}{\bibinfo{person}{Jean-Marc Valin}, \bibinfo{person}{Gregory
  Maxwell}, \bibinfo{person}{Timothy~B Terriberry}, {and} \bibinfo{person}{Koen
  Vos}.} \bibinfo{year}{2016}\natexlab{}.
\newblock \showarticletitle{High-quality, low-delay music coding in the opus
  codec}.
\newblock \bibinfo{journal}{\emph{arXiv preprint arXiv:1602.04845}}
  (\bibinfo{year}{2016}).
\newblock


\bibitem[\protect\citeauthoryear{Valin, Vos, and Terriberry}{Valin
  et~al\mbox{.}}{2012}]%
        {valin2012definition}
\bibfield{author}{\bibinfo{person}{Jean-Marc Valin}, \bibinfo{person}{Koen
  Vos}, {and} \bibinfo{person}{Timothy Terriberry}.}
  \bibinfo{year}{2012}\natexlab{}.
\newblock \bibinfo{booktitle}{\emph{Definition of the Opus audio codec}}.
\newblock \bibinfo{type}{{T}echnical {R}eport}.
\newblock


\bibitem[\protect\citeauthoryear{Vodafone}{Vodafone}{2019}]%
        {5gdigvodafone}
\bibfield{author}{\bibinfo{person}{Vodafone}.} \bibinfo{year}{2019}\natexlab{}.
\newblock \bibinfo{title}{5GDig: The winners -- from Skype for surgeons to AR}.
\newblock
  \bibinfo{howpublished}{\url{https://www.vodafone.com/news/technology/5gdig-winners-2019-supponor-ar-surgeonmate}}.
\newblock


\bibitem[\protect\citeauthoryear{Vos, Jensen, and Soerensen}{Vos
  et~al\mbox{.}}{2010}]%
        {vos2010silk}
\bibfield{author}{\bibinfo{person}{Koen Vos}, \bibinfo{person}{Soeren Jensen},
  {and} \bibinfo{person}{Karsten Soerensen}.} \bibinfo{year}{2010}\natexlab{}.
\newblock \showarticletitle{SILK speech codec}.
\newblock \bibinfo{journal}{\emph{IETF draft}}  \bibinfo{volume}{30}
  (\bibinfo{year}{2010}).
\newblock


\bibitem[\protect\citeauthoryear{Zhang and Sabuncu}{Zhang and Sabuncu}{2018}]%
        {zhang2018generalized}
\bibfield{author}{\bibinfo{person}{Zhilu Zhang} {and} \bibinfo{person}{Mert
  Sabuncu}.} \bibinfo{year}{2018}\natexlab{}.
\newblock \showarticletitle{Generalized cross entropy loss for training deep
  neural networks with noisy labels}. In \bibinfo{booktitle}{\emph{Advances in
  neural information processing systems}}. \bibinfo{pages}{8778--8788}.
\newblock


\end{thebibliography}

\end{document}